\documentclass[twocolumn]{aastex631}
\usepackage{amsmath}
\usepackage{enumerate}
\usepackage{enumitem}   
\usepackage{color, colortbl}
\usepackage{subfigure}
\usepackage{color}
\usepackage{courier}
\usepackage{mathtools}
\usepackage{threeparttable}
\usepackage{enumitem}
\usepackage{rotating}
\usepackage{tabularx}
\usepackage{tabularx}
\usepackage{rotating}
\usepackage{CJK}
\usepackage{multirow}
\usepackage[]{threeparttable}

\newcommand\clearrow{\global\let\rowmac\relax}

\newcommand{\Mstar}{$M_{\ast}$}
\newcommand{\Sigstar}{$\Sigma_{\ast}$}
\newcommand{\fgas}{$f_\mathrm{gas}$}
\newcommand{\Sigsfr}{$\Sigma_\mathrm{SFR}$}
\newcommand{\Sigssfr}{$\Sigma_\mathrm{sSFR}$}
\newcommand{\Sigmol}{$\Sigma_\mathrm{H2}$}
\newcommand{\highsf}{SF$_{h}$}
\newcommand{\midsf}{SF$_{m}$}
\newcommand{\lowsf}{SF$_{l}$}
\newcommand{\Reff}{$R_\mathrm{e}$}

\received{}
\revised{}
\accepted{}
\submitjournal{}

\begin{document}

\title{The ALMaQUEST Survey XIII: Understanding radial trends in star formation quenching via the relative roles of gas availability and star formation efficiency}

\begin{abstract} 
Star formation quenching is one of the key processes that shape the evolution of galaxies. 
In this study, we investigate the changes in molecular gas and star formation properties as galaxies transit from the star-forming main sequence to the passive regime. 
Our analysis reveals that as galaxies move away from the main sequence towards the green valley the radial profile of specific star formation rate surface density (\Sigssfr) is suppressed compared with main sequence galaxies out to a galactocentric radius of 1.5\Reff ($\sim$ 7 kpc for our sample).
By combining radial profiles of gas fraction (\fgas) and star formation efficiency (SFE), we can discern the underlying mechanism that determines \Sigssfr\ at different galactocentric radii.
Analysis of relative contributions of \fgas\ and SFE to \Sigssfr\ uncovers a diverse range of quenching modes. 
Star formation in approximately half of our quenching galaxies is
 primarily driven by a single mode (i.e. either \fgas\ or SFE),  or a combination of both. 
A collective analysis of all galaxies reveals that the reduction in star formation within the  central regions ($R$~$<$~0.5~\Reff) is primarily attributable to a decrease in SFE. Conversely, in the  disk regions ($R$~$>$~0.5~\Reff), both \fgas\ and SFE contribute to the suppression of star formation. 
Our findings suggest that multiple quenching mechanisms may be at play in our sample galaxies, and even within a single galaxy.
We also compare our observational outcomes with those from galaxy  simulations and discuss the implications of our data.

\end{abstract} 

\begin{CJK}{UTF8}{bkai}
\author{Hsi-An Pan (潘璽安)}
\affiliation{Department of Physics, Tamkang University, No.151, Yingzhuan Road, Tamsui District, New Taipei City 251301, Taiwan}

\author{Lihwai Lin}
\affiliation{Institute of Astronomy and Astrophysics, Academia Sinica, Taipei 10617, Taiwan}

\author{Sara L. Ellison}
\affiliation{Department of Physics and Astronomy, University of Victoria, Finnerty Road, Victoria, British Columbia V8P 1A1, Canada}

\author{Mallory D. Thorp}
\affiliation{Department of Physics and Astronomy, University of Victoria, Finnerty Road, Victoria, British Columbia V8P 1A1, Canada}
\affiliation{Argelander-Institut f{\"u}r Astronomie, Universit{\"a}t Bonn, Auf dem H{\"u}gel 71, 53121 Bonn, Germany}

\author{Sebasti\'{a}n F. S\'{a}nchez}
\affil{Instituto de Astronom\'{i}a, Universidad Nacional Aut\'{o}noma de M\'{e}xico, A. P. 70-264, C.P. 04510, M\'{e}xico, D.F., Mexico}

\author{Asa F. L. Bluck}
\affiliation{Department of Physics, Florida International University, 11200 SW 8th Street, Miami, FL 33199, USA}

\author{Francesco Belfiore}
\affiliation{INAF— Osservatorio Astrofisico di Arcetri, Largo E. Fermi 5, I-50125,	Florence, Italy}

\author{Joanna M. Piotrowska}
\affiliation{Cahill Center for Astrophysics, California Institute of Technology, 1216 East California Boulevard, Pasadena, CA 91125, USA}

\author{Jillian M. Scudder}
\affiliation{Department of Physics \& Astronomy, Oberlin College, Oberlin, OH, 44074, USA}

\author{William M. Baker}
\affiliation{Kavli Institute for Cosmology, University of Cambridge, Madingley Road, Cambridge, CB3 0HA, UK}
\affiliation{Cavendish Laboratory—Astrophysics Group, University of Cambridge, 19 JJ Thomson Avenue, Cambridge, CB3 0HE, UK}

\section{Introduction}
Galaxies are typically classified into two main populations on the global star formation rate (SFR) versus stellar mass (\Mstar) diagram: the star-forming main sequence and the passive population \citep{Str21,Bri04,Elb07,Noe07}. 
The star-forming main sequence consists of mostly disk galaxies with higher specific star formation rate (sSFR = SFR/\Mstar, where \Mstar\ is global stellar mass of a galaxy), while the passive population is predominantly composed of quiescent spheroidal galaxies with lower sSFR \citep{Sch14}. 
However, there is an intermediate population that is sparsely populated between these two regimes, known as the green valley \citep{Wyd07,Mar07,Sal14}. 
Green valley galaxies represent a transitional population that is likely undergoing a shift from the  star-forming phase to the  quiescent phase. 
Therefore, studying the physical properties of green valley galaxies is crucial for gaining insights into the processes of star formation quenching and galaxy evolution.

On a spatially-resolved scale, previous works have shown that the differences in global sSFR between star-forming, green valley, and quiescent galaxies is driven by variations in the fraction of ``retired'' regions that have ceased to form new stars, with red galaxies exhibiting the highest fraction of retired spaxels, and vice versa \citep{Hsi17,Pan18a,Can19,Lin22}. 
Since star formation primarily occurs in molecular gas clouds, the presence of retired regions is expected to be linked to the changes in the properties of molecular gas. 
Specifically, the decline in star formation activity can be attributed to a depletion of molecular gas and/or a reduction in its efficiency in forming stars, which could result from factors such as inefficient cooling or turbulent pressure support \citep{Mck07}. 
These scenarios correspond to the changes in two observable parameters: the gas fraction (\fgas), defined as the molecular gas mass of a galaxy relative to its stellar mass ($M_\mathrm{H_{2}}$/\Mstar), and the star formation efficiency (SFE), which quantifies the conversion of gas into stars as the number of stars formed per unit time (SFR/$M_\mathrm{H_{2}}$). 
Previous global studies of molecular gas have demonstrated that star formation quenching is associated with changes in both global \fgas\ and SFE \citep[e.g.,][]{Sar14,Sai16,Col20,Pio20,Wyl22}, but there is no definitive conclusion regarding the relative importance of one factor over the other.

A crucial step towards understanding the process of star formation quenching is the investigation of the spatially-resolved molecular gas properties of galaxies. The ALMaQUEST survey (ALMA-MaNGA Quenching and Star-Formation Survey;  \citealt{Lin20})  was specifically designed to study the resolved molecular gas properties of various types of galaxies, including those above \citep{Ell20a} and below \citep{Lin22} the global star-forming main sequence, as well as galaxy mergers \citep{Tho22}. 

Our most recent studies have unveiled important insights into the molecular gas properties of galaxies. 
Specifically, we have found that retired regions within galaxies generally exhibit lower \fgas\ and/or SFE compared to active star-forming regions \citep{Ell21b,Lin22}. 
Interestingly, we have also observed that even the star-forming regions within green-valley galaxies, which are in a transitional state, show lower sSFR, SFE, and \fgas\ compared to  star-forming regions of the star-forming main sequence (\citealt{Lin22}, see also \citealt{Bro20}  for the similar results, but from different observations). 
This suggests that the overall reduction in sSFR in green-valley galaxies is a result of not only an increased fraction of retired regions but also by a decrease in sSFR within their active star-forming regions (\citealt{Lin22}, see also \citealt{Ell18} and \citealt{Bel18}).
The suppressed sSFR, \fgas, and SFE in green-valley galaxies naturally gives rise to a range of resolved scaling relations that deviate from those observed in normal star-forming galaxies
 (\citealt{Ell21a,Bak22,Lin22,Bak23,Ell23}; see also \citealt{Gar23} for a discussion based on non-ALMaQUEST observations).

On the other hand, the chosen analytical approach also plays a pivotal role in interpreting the data. 
For instance, our analysis of the ALMaQUEST merger sample (cf. Section \ref{sec_almaquest}) reveals that when all spaxels are considered together, one might conclude that approximately half of the interaction-triggered star formation is driven by \fgas, and the remaining half by SFE \citep{Tho22}. 
However, when examined on a per-galaxy basis, it becomes evident that some galaxies are entirely driven by \fgas, while others are exclusively powered by SFE. 
Therefore, combining all data may lead to a loss of information.

While our previous ALMaQUEST studies that focus on star formation quenching \citep{Ell21b,Lin22} have primarily considered the integral behaviors of the entire sample, 
this work specifically focuses on the continuous changes in star formation and molecular gas properties as a function of a galaxy's distance to the star-forming main sequence. 
Furthermore, we investigate the relative roles of \fgas\ and SFE within individual galaxies, an aspect that has not been extensively addressed thus far. 
Specifically, we aim to answer questions regarding 1) \emph{how the radial profiles of star formation and molecular gas properties change as a function of a galaxy's distance to the star-forming main sequence}, and 2) \emph{whether there is a preference for a specific quenching driver (\fgas\ or SFE) at different galactocentric radii}.
Finally, we will compare our observational findings with those obtained from galaxy simulations in the literature.

The paper is structured as follows: Section \ref{sec_data} provides a  description of the multiwavelength data employed in this study. 
The main findings are presented in Section \ref{sec_results}, including the radial profiles of star formation and molecular gas properties in Section \ref{sec_rad} and the relative importance of \fgas\ and SFE in Section \ref{sec_relative_importance}.
We   discuss the implications of our results and compare our observational results with those from galaxy simulations  in Section \ref{sec_discussion}.
Section \ref{sec_summary} provides a  summary of the results obtained in this work.
Throughout this paper, we adopt the following cosmological parameters: $H_\mathrm{0}$ = 70 km s$^{-1}$ Mpc$^{-1}$, $\Omega_\mathrm{m}$ = 0.3, and $\Omega_\mathrm{\Lambda}$ = 0.7. 
A Salpeter initial mass function \citep{Sal55} is assumed in this work.

\section{Data and Observations}
\label{sec_data}
\subsection{The ALMaQUEST Survey}
\label{sec_almaquest}
The main ALMaQUEST survey \citep{Lin20}\footnote{http://arc.phys.uvic.ca/~almaquest/} aims to investigate the molecular gas properties of 46 galaxies  selected from the MaNGA (Mapping Nearby Galaxies at APO) integral-field unit (IFU) spectroscopic  survey \citep{Bun15}.
The 46 galaxies consist of 34 non-starburst and 12 starburst galaxies.
The survey focuses on galaxies spanning a wide range of log(sSFR/yr$^{-1}$) from $-$9.4 to $-$12.2, covering the starburst to green valley regime below the global star-forming main sequence on the SFR versus \Mstar\ plane. 
The ALMaQUEST sample have been expanded to include 20 galaxy mergers  \citep{Tho22,Ell23}, which are not included in this work.

The stellar mass of the ALMaQUEST main sample spans from log(\Mstar/M$_{\odot}$) $=$ 10.2 to 11.6.
While the majority of the galaxies are selected based on their global \Mstar\ and sSFR, an additional criterion is applied to select 12 starburst galaxies. 
These galaxies are required to exhibit an elevated  \Sigsfr\ relative to the control spaxels within 0.5\Reff\ by at least 50\% \citep[see][for details]{Ell20a}.
These control spaxels are characterized by their similarity in \Sigstar\ and location within a comparable galactic radius to the spaxel in question, and belong to galaxies with similar global \Mstar. 
Since the selection of starburst galaxies in \cite{Ell20a} was based on the \emph{relative} profile, rather than an \emph{absolute} value, these galaxies do not necessarily exhibit higher sSFR, both locally and globally, compared to other galaxies in the ALMaQUEST sample, although a majority of them indeed do.
It is important to note that other galaxy properties and environmental factors are not considered in the sample selection for the ALMaQUEST survey.

\subsection{Star Formation Tracers: MaNGA}
\label{sec_manga}
The MaNGA data used in this study are obtained from the MaNGA DR15 \texttt{Pipe3D} value-added products \citep{San18}. 
These \texttt{Pipe3D} products provide both global properties of galaxies integrated over the MaNGA bundle field-of-view, as well as spaxel-based measurements of local stellar mass and emission-line fluxes.
The stellar mass is derived based on the best-fit stellar population model that describes the stellar continuum of a given spectrum. 
The best-fit stellar continuum is then subtracted from the  spectrum to obtain the emission-line measurements. 
To account for the effect of dust extinction, we apply a correction to the emission line fluxes using the Balmer decrement computed at each spaxel. 
The correction is based on a Milky Way extinction curve with a standard reddening parameter $R_{V}$ of 3.1 \citep{Car89}.

In this study, we adopt H$\alpha$ emission as a tracer of ongoing star formation activity. 
The extinction-corrected H$\alpha$ luminosity is converted to the SFR using the calibration provided by \cite{Ken98}.
To calculate the SFR and stellar mass surface densities (\Sigsfr\ and \Sigstar) for each spaxel, we utilize the spaxel-based stellar mass and SFR measurements from \texttt{Pipe3D}, taking into account the physical area of the spaxel and applying an inclination correction.

However, it is important to note that H$\alpha$ emission can arise from sources other than young stars, such as AGN and old stellar populations   \citep{Kew06,Yan12}.
To classify the spaxels based on the powering sources, we follow the method outlined in our previous papers \citep{Blu20a,Lin22}.
In summary, we first utilize the Baldwin--Phillips--Terlevich (BPT) diagnostics, a commonly used tool for distinguishing between different sources of H$\alpha$ emission.
Specifically, we apply the Kewley and Kauffmann  star-forming cut based on the [\ion{O}{3}]$\lambda$5007/H$\beta$ versus [\ion{N}{2}]$\lambda$6584/H$\alpha$ line ratios \citep{Kau03,Kew06} to classify spaxels into  star formation, composite, or AGN.
Our BPT analysis is limited to spaxels with a signal-to-noise ratio (S/N) greater than 2 for all required emission lines.
\emph{Star-forming spaxels} are defined as those satisfying both the BPT star-forming criteria and an H$\alpha$ equivalent width (EW) greater than 6\AA\  \citep{San20,San21}.
H$\alpha$ luminosity is used to estimate the  \Sigsfr\ for star-forming spaxels.

\begin{figure}
	\centering
	\includegraphics[scale=0.53]{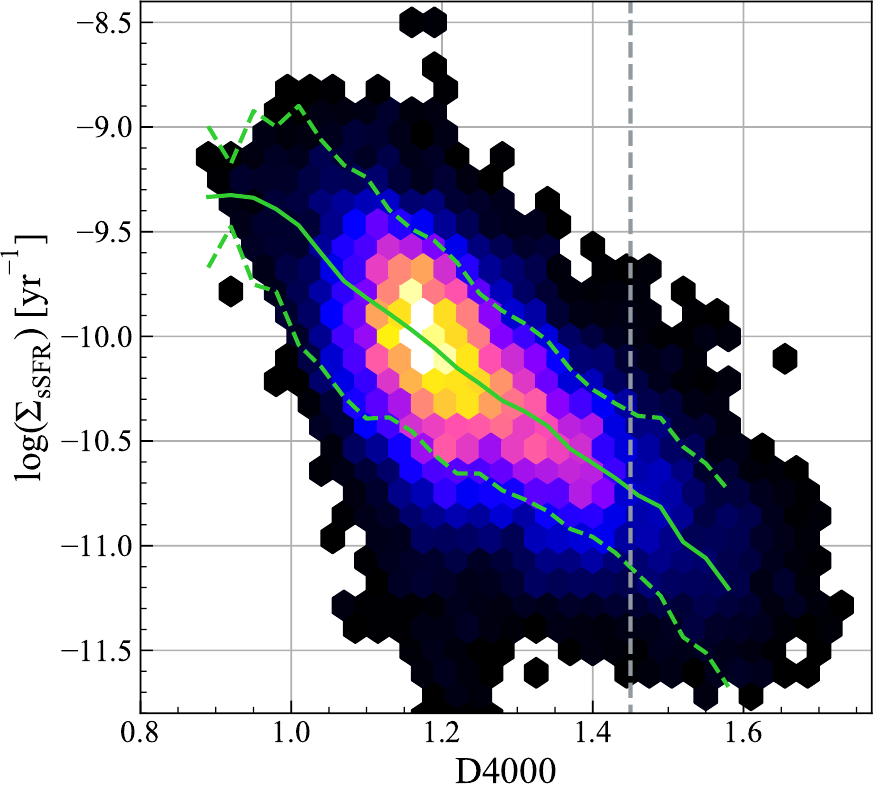}
	\caption{D4000 and \Sigssfr\ relation for the star-forming spaxels in the ALMaQUEST sample. The color scale indicate the density of spaxel at each D4000 and \Sigssfr\ bins.  The median \Sigssfr\ value  in each D4000 bin is  as green solid curve. The standard deviation of \Sigssfr\ within each bin is shown as a green dashed lines. 
	Spaxels with D4000 $>$1.45 (marked by a vertical gray line) are fully quenched and therefore the sSFR-D4000 relation is no longer viable \citep{Blu20a}. Therefore, D4000-based \Sigssfr\ is computed for AGN-contaminated and retired spaxels only when their D4000 is lower than 1.45. The fully quenched spaxels (D4000 $>$ 1.45) are artificially assigned to a fixed value of log(\Sigssfr/yr$^{-1}$) $=$ $-$12.
	}  
	\label{fig_D4000_SFR}
\end{figure}

We also compute  \Sigsfr\  for two categories of non-star-forming spaxels.
The \emph{AGN-contaminated spaxels} are those in the AGN and composite regimes on the BPT diagram.
The \emph{retired spaxels}  where the star formation is largely suppressed are identified as those having S/N $>$ 2 and  EW $<$ 3 \AA\ in H$\alpha$ \citep{Lin22}.
We estimate \Sigsfr\ for spaxels that are not dominated by active star formation (either AGN-contaminated or retired spaxels)  using the relationship between \Sigsfr\ and 4000 \AA\ break (D4000) as employed in previous MaNGA and ALMaQUEST papers \citep[e.g.,][]{Spi18,Wan19,Blu20a,Tho22}.
In Figure \ref{fig_D4000_SFR}, we show the method used to estimate \Sigsfr\ for these spaxels. 
We first construct the \Sigssfr-D4000 relation for our sample using star-forming spaxels from the 34 non-starburst star forming galaxies. 
The median \Sigssfr\ value in each D4000 bin is calculated and represented by the green solid curve. 
For any spaxel that does not meet our star-forming criteria  based on the BPT and EW cuts, we assign a \Sigssfr\ value based on its D4000 and the corresponding median curve. 
Then, we convert the \Sigssfr\ to \Sigsfr\ by multiplying it with the \Sigstar\ value of these spaxels.
The empirical \Sigssfr-D4000 relation shown in Figure \ref{fig_D4000_SFR} is consistent with that derived from the full MaNGA sample \citep[][]{Blu20a,Tho22}.

Spaxels with D4000 $>$ 1.45 are generally considered quenched, and thus the sSFR-D4000 relation is no longer applicable \citep{Wan18,Blu20a}. 
Therefore, we only utilize the \Sigssfr-D4000 calibration for spaxels with D4000 $<$ 1.45. 
This means that the \Sigssfr-D4000 calibration limits the analysis to the suppressed (D4000 $<$ 1.45), but not fully quenched (D4000 $>$ 1.45) regions  (see the discussion in \citealt{Wan18} and \citealt{Blu20a}).
Note that since the \Sigsfr\ values of the   AGN-contaminated and retired spaxels are inferred based on their D4000 values, some accuracy is traded for completeness. 
Thus, the \Sigsfr\ values of these spaxels should be interpreted with caution.

Finally, it is  important to utilize the full sample of spaxels in order to probe quenching \citep{Blu20a}.
Therefore, we define \emph{fully-quenched spaxels} as those having S/N $<$ 2 in H$\alpha$ or D4000 $>$  1.45. 
In the examined dataset, star-forming spaxels with a D4000 value of 1.45 span a range from $-$11.7 to $-$9.5 in log(\Sigssfr/yr$^{-1}$), with a median value of -10.7.
Therefore, the fully-quenched spaxels  are artificially assigned to a fixed value of log(\Sigssfr/yr$^{-1}$) $=$ $-$12 (i.e., just below $-$11.7 as an upper limit for \Sigssfr), following \cite{Blu20a}.
Our spaxel classification that includes star-forming, AGN-contaminated, retired, and fully-quenched spaxels ensures that $\sim$ 98\% of spaxels have a valid \Sigsfr\ value.
This allows for the construction of a continuous  star formation profile as a function of  galactocentric radius  for all the galaxies in this work.
Nevertheless, we have verified that our conclusions remain qualitatively consistent even when excluding the fully-quenched spaxels, which were assigned log(\Sigssfr/yr$^{-1}$) $=$ $-$12  manually.

\subsection{Molecular Gas Observations: ALMA}
\label{sec_alma}
The ALMA observations at $^{12}$CO ($J$ $=$ 1 $\rightarrow$ 0; 115.271 GHz) were designed to match the point-spread function (2.5$\arcsec$) of the MaNGA observations, allowing for a simultaneous study of spatially-resolved star formation and molecular gas properties at the same physical scale  (PI: L. Lin -- 2015.1.01225. S, 2017.1.01093.S, 2018.1.00558.S; PI: S. Ellison -- 2018.1.00541.S). 
Detailed information regarding ALMA data reduction and imaging can be found in the survey paper by \cite{Lin20}.

In this study, the molecular mass surface density (\Sigmol) is calculated from the $^{12}$CO ($J$ $=$ 1 $\rightarrow$ 0) luminosity using a constant Milky Way conversion factor ($\alpha_\mathrm{CO}$) of 4.35 M$_{\odot}$ (K km s$^{-1}$ pc$^{2}$)$^{-1}$ \citep{Bol13}. 
The value of $\alpha_\mathrm{CO}$ is not universal and can vary with factors such as metallicity, cosmic ray density, and UV radiation field \citep[see the review by][]{Bol13}. 
The assumption of a constant $\alpha_\mathrm{CO}$ has been examined in previous ALMaQUEST papers \citep{Ell21a,Lin20,Ell23}. 
On average, the difference in the global molecular gas mass ($M_\mathrm{H_2}$) obtained using a metallicity-dependent $\alpha_\mathrm{CO}$ compared to the original $M_\mathrm{H_2}$ is approximately 0.1 dex. 
Given the small difference, a constant $\alpha_\mathrm{CO}$ is adopted in this work to maintain consistency with previous ALMaQUEST papers.
However, it is worth noting that although the radial profiles of $\alpha_\mathrm{CO}$ are generally flat within a given galaxy at kpc scale, galaxies often exhibit a decrease in $\alpha_\mathrm{CO}$ within the inner 1 kpc \citep{San13}, which corresponds to (or is slightly smaller than) the spatial scale that our ALMA and MaNGA observations can resolve.

To recover the CO non-detection spaxel for spaxel completeness, a 3$\sigma$ upper limit is assigned to spaxels with S/N $<$ 2 in CO integrated intensity map. 
The 3$\sigma$ upper limits of flux are calculated with $3\sigma\times{\sqrt{\delta v \Delta V}}$, where $\sigma$ is the rms noise from the spectral line data cube, $\delta v$~$=$~11~km~s$^{-1}$ is the velocity resolution of the data cube, and $\Delta V$ is the assumed 40~km~s$^{-1}$ line width for a bulk of molecular cloud observed at kpc resolution. 
Two methods were employed for measuring the local line width: pixel-by-pixel Gaussian fitting of the data cubes and the CASA task \texttt{immoment} (moment 2), which represents intensity-weighted line width around the mean without making any assumptions about the line profile.
Both methods show a median line width of $\sim$40~km~s$^{-1}$.
However, it is important to note that these estimates inherently rely on the spaxels that have a CO detection.

\subsection{Sample Classification}
\label{sec_sample_classification}
The primary goal of this study is to investigate the continuous variations in gas and star formation properties as galaxies transit away from the star-forming main sequence. 
To achieve this, we classify the 34 non-starburst galaxies into three groups based on their deviation from the global star-forming main sequence ($\Delta$MS).
The 12 starburst galaxies (which are included in our sample, but treated as a separate class, see below) are also on or slightly above the main sequence, exhibiting enhanced SFR in the centers by design.

\begin{figure}
	\centering
	\includegraphics[scale=0.53]{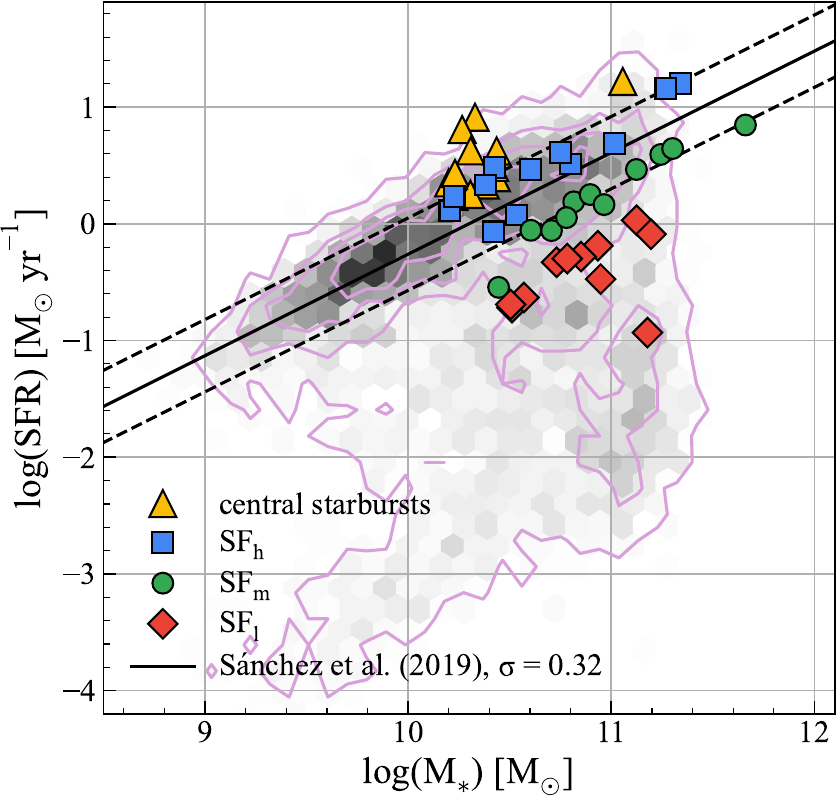}
	\caption{Distribution of our sample on the global stellar mass (\Mstar) versus star formation rate (SFR) plane. The yellow triangles represent the central starburst galaxies (cSB) in the ALMaQUEST survey.  The blue squares, green circles and red diamonds correspond to the \highsf\ (sSFR (log(sSFR/yr$^{-1}$) $>$ $-$10.5), \midsf\ ($-$11.0 $<$ log(sSFR/yr$^{-1}$) $<$ $-$10.5), and \lowsf\ ((log(sSFR/yr$^{-1}$) $<$ $-$11.0) galaxies  in this work, respectively. They are classified based on  the absolute value of global sSFR. 
	The solid black line marks the star-forming main sequence (Equation \ref{eq_sfms})  estimated by  \cite{San19} for the MaNGA sample.
	The scatter of the main sequence is $\sim$ 0.32 dex (dashed lines). The contours represent the \Mstar\ and SFR distribution of $\sim$ 5000 galaxies included in data release SDSS DR15 measured by Pipe3D.
	}  
	\label{fig_sample}
\end{figure}

Figure \ref{fig_sample} illustrates the global SFR-\Mstar\ relation of our sample. 
The background contours and gray scale show the galaxies from the  MaNGA (SDSS DR15) sample. 
The solid and dashed black lines represent the global star-forming main sequence and its typical scatter of $\pm$0.32 dex for the MaNGA sample, as determined by \cite{San19} as 
\begin{equation}
	\log(\mathrm{SFR_{H\alpha}})= (-8.96\pm0.23) + (0.87\pm0.02) \log(M_{\star}),
\label{eq_sfms}
\end{equation}
where  the SFR is in units of M$_{\sun}$~yr$^{-1}$ and \Mstar\ is in units of M$_{\sun}$.

In the work presented here, we have categorized the sample into three distinct classes based on their sSFR:  those with relatively high global sSFR (log(sSFR/yr$^{-1}$) $>$ $-$10.5; blue squares in Figure \ref{fig_sample}), moderate sSFR ($-$11.0 $<$ log(sSFR/yr$^{-1}$) $<$ $-$10.5; green circles), and low sSFR (log(sSFR/yr$^{-1}$) $<$ $-$11.0; rad diamonds), which we hereafter refer to as the \highsf, \midsf, and \lowsf\ samples, respectively. 
Broadly speaking, \highsf\ galaxies are the main-sequence, normal star-forming galaxies, \midsf\ galaxies are located  at the lower end or just below the main-sequence, and \lowsf\ galaxies reside below the main-sequence.
For the 34 non central starburst galaxies, there are a total of 12 galaxies classified as \highsf, 11 galaxies classified as \midsf, and 11 galaxies classified as \lowsf.
Although the sample classification is defined by the absolute value of global sSFR in this work, it is correlated with the distance to the star-forming main sequence ($\Delta$MS), which has been used in other studies to distinguish star-forming and quenched galaxies \citep[e.g.,][]{Ell18}.
Our sSFR boundaries in absolute unit correspond to $\Delta$MS of $-$0.2 $<$ $\Delta$MS $<$ $+$0.4 dex for \highsf, $-$0.7 $<$ $\Delta$MS $<$ $-$0.2 dex for \midsf, and $-$1.7 $<$ $\Delta$MS $<$ $-$0.7 dex for \lowsf, where the $\Delta$MS is calculated as the log difference between the SFR and the star-forming main sequence in Equation \ref{eq_sfms} for given \Mstar.
Table \ref{tab_sample_class} provides an overview of the sample classification and the corresponding range of global sSFR and $\Delta$MS.

The central starburst galaxies (cSB) are shown as yellow triangles in Figure \ref{fig_sample}.
cSB are all lying above the line of star-forming main sequence, with $\Delta$MS of 0.04 -- 0.90 dex.
In the subsequent analysis, we will focus on characterizing the spatial distribution of various gas and star formation properties using radial analysis. 
As such, the 12 cSBs that were specifically selected to exhibit centrally peaked starbursts \citep{Ell20a}  are treated as their own class.
They are presented where applicable for the purpose of comparison.

\begin{table}[]
	\centering
	\begin{tabular}{cccc}
		\hline
		&  log(sSFR){[}yr$^{-1}${]} & $\Delta$MS {[}dex{]} & \# of galaxies \\
		\hline
		& \multicolumn{3}{c}{star-forming main sequence}      
	                    \\
	    		\hline
		\highsf & $>$ $-$10.5                     & $-$0.2 to $+$0.4     & 12             \\
				\hline
		& \multicolumn{3}{c}{quenching galaxies}  
				                             \\
				                             \hline   
		\midsf & $-$11.0 to $-$10.5              & $-$0.7 to $-$0.2     & 11             \\
		\lowsf & $<$ $-$11.0                     & $-$1.7 to $-$0.7     & 11  \\   
				\hline       
	\end{tabular}
	\caption{Sample classification. The sample classification is defined by the absolute value of global sSFR (second left column), it is correlated with the distance to the star-forming main sequence ($\Delta$MS; second right column) which has been used in other studies to distinguish star-forming and quenched galaxies. The number of galaxies in each category is presented in the right column.}
	\label{tab_sample_class}
\end{table}

\section{Results}
\label{sec_results}

\subsection{Radial Profiles of Star Formation and Molecular Gas Properties}
\label{sec_rad}

\subsubsection{Radial Profiles of  \Sigstar, \Sigmol\, and \Sigsfr}
\label{sec_radial_raw}
\begin{figure*}
	\centering
	\includegraphics[scale=0.53]{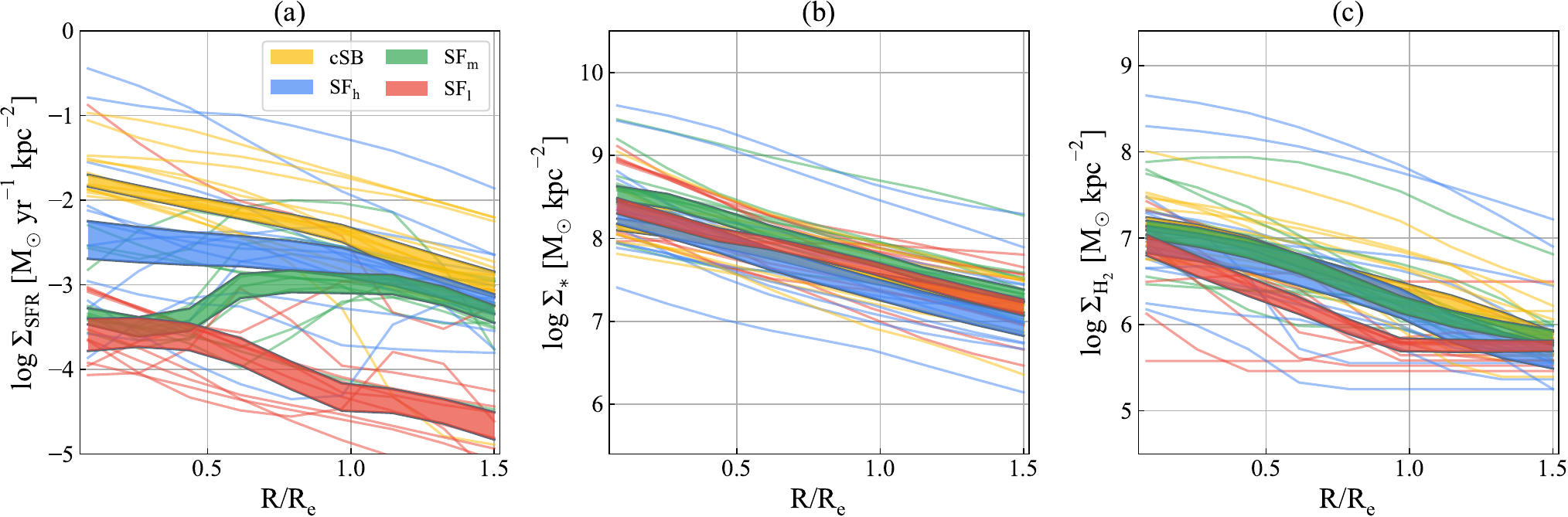}
	\caption{Panel (a) -- (c): Radial profiles of \Sigsfr, \Sigstar, and \Sigmol, respectively. The thin lines show the the radial properties of each individual galaxy, with color indicating the population of the galaxies (yellow: central starburst galaxies cSB, blue: main-sequence galaxies \highsf, green: galaxies just below the main sequence \midsf, and red: galaxies well below the main sequence \lowsf). The colored thick lines denote the median profiles of the properties for galaxies  in the corresponding category.  The height of the shaded regions  represents the standard error of the mean at each radial bin.
	}  
	\label{fig_radial_phyprops}
\end{figure*}

We begin by analyzing the galaxy radial profiles in the ALMaQUEST sample. 
Figures \ref{fig_radial_phyprops}(a)-(c) present the radial profiles of \Sigsfr, \Sigstar, and \Sigmol, respectively. 
The subsets, namely \highsf, \midsf, \lowsf, and cSB, are denoted by blue, green, red, and yellow colors, respectively.
The radial profile is sampled with a 0.18~$R_\mathrm{e}$ ($\sim$ 0.8 kpc) steps within 1.5~$R_\mathrm{e}$ ($\sim$ 7 kpc at redshift 0.03).
The inclination angle, along with the galaxy's position angle, is considered to deproject the geometries of the galaxies.
The radial quantities are determined by calculating the median value\footnote{In this work, we choose median instead of mean so that the result suffers less from some extreme outliers.} of the star formation and molecular gas properties within each deprojected annulus.
Radial profiles are constructed using all spaxels with valid values, regardless of whether these values were measured, inferred, or represented as upper limits (see Section \ref{sec_manga} and \ref{sec_alma}).
As mentioned earlier, this is to probe the quenched and gas-deficient regions.
For each individual galaxy, we display the radial properties using thin lines. 
The median profiles of each subset, which are obtained by taking the median values from the individual profiles, are shown using shaded colored regions. 
The height of the shaded regions  represents the standard error of the mean at each radial bin.
Note again that the cSB subset should be considered as a distinct population since the sample was selected based on their radial profiles of local sSFR by design (cf. Section \ref{sec_almaquest}).

The radial profiles of \Sigsfr\ exhibit a ranking across the explored radial range, with a consistent increase  in magnitude from \lowsf\ to \midsf, \highsf, and to cSB galaxies at almost all radii (Figure \ref{fig_radial_phyprops}(a)). 
Notably, the differences between \highsf, \midsf, and cSB populations are more pronounced in the inner region ($<$ 0.5 \Reff) of galaxies compared to the outer region ($>$ 1.0 \Reff), consistent with the finding in previous work with larger sample size \citep[e.g.,][]{Ell18,Wan19,Blu20a}.
Conversely, the \Sigsfr\ of \lowsf\ galaxies is significantly lower than that of the other populations at all galactocentric radii.
Several \highsf\ galaxies exhibit a \Sigsfr\ higher than that of cSB galaxies. 
However, note that when normalized by \Sigstar\ (Figure \ref{fig_radial_ratios}(a)), these galaxies typically show a lower \Sigssfr\  compared to cSB galaxies.

\begin{figure*}
	\centering
	\includegraphics[scale=0.53]{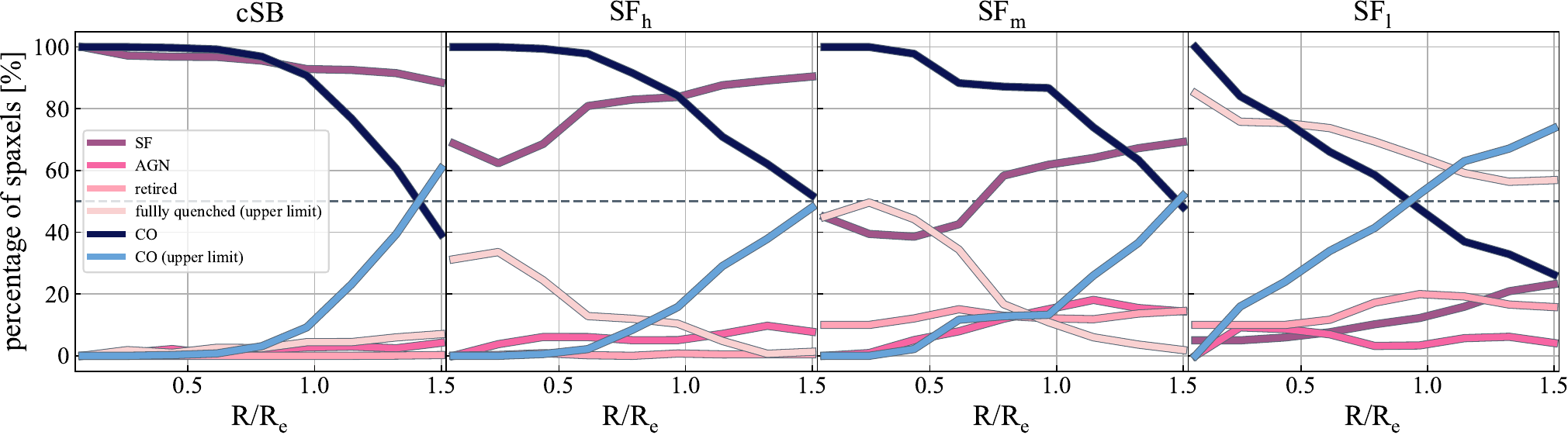}
	\caption{Panel(a) -- (d): Fraction of spaxels classified as star-forming (in purple), AGN-contaminated (in magenta), retired (in pink), and fully-quenched (in light pink) spaxels at various radial bins for cSB, \highsf, \midsf, and \lowsf\ galaxies, respectively.
	The dark blue and blue curves represent the fraction of spaxels with and without significant CO detection.
	The horizontal line represents a spaxel percentage of 50\%. Caution should be taken when interpreting the radial profiles in Figure \ref{fig_radial_phyprops} and \ref{fig_radial_ratios}, particularly when the fraction of spaxels with upper limits exceeds 50\%.
	}  
	\label{fig_radial_spxtype}
\end{figure*}

\begin{figure*}
	\centering
	\includegraphics[scale=0.53]{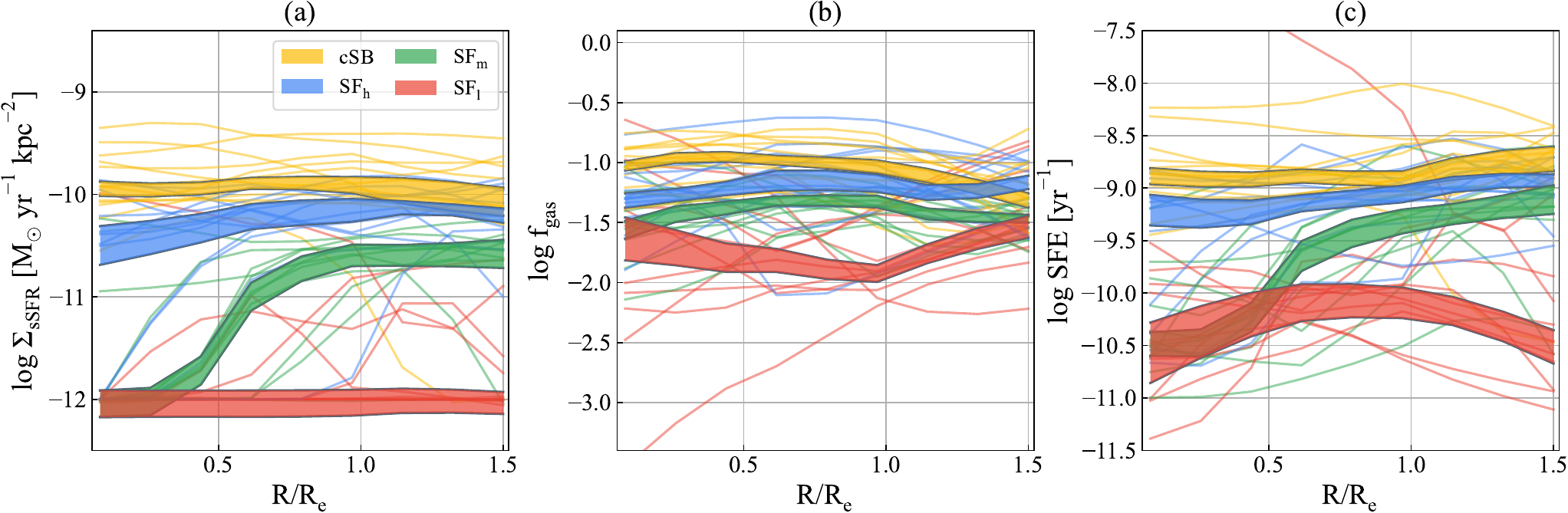}
	\caption{Panel (a) -- (c): Radial profiles of \Sigssfr, \fgas, and SFE, respectively. The thin lines show the the radial properties of each individual galaxy, with color indicating the population of the galaxies (yellow: central starburst galaxies cSB, blue: main-sequence galaxies \highsf, green: galaxies just below the main sequence \midsf, and red: galaxies well below the main sequence \lowsf). The colored thick lines denote the median profiles of the properties for galaxies  in the corresponding category. The height of the shaded regions  represents the standard error of the mean at each radial bin. 
	}  
	\label{fig_radial_ratios}
\end{figure*}

Due to the trade-off between \Sigsfr\ accuracy and completeness, as discussed in Section \ref{sec_manga}, it is important to interpret Figure \ref{fig_radial_phyprops}(a) with caution.
Figure \ref{fig_radial_spxtype} shows the radial profiles of the percentage of different type of spaxels.
All the available spaxels from each individual galaxy class are combined to generate the plot.
Results for cSB, \highsf, \midsf, and \lowsf\ galaxies are presented in panel (a) to (d), respectively.
As shown in the figure, there is a significant decline in the fraction of star-forming spaxels (purple curve) when transitioning from cSB to \lowsf\ galaxies. 
Within \lowsf\ galaxies (panel (d)), fully-quenched regions (light pink) dominate, constituting over 50\% of the regions at all radii. 
In these regions, the log(\Sigssfr/yr$^{-1}$) is manually set to $-$12, and therefore their \Sigsfr\ values depend on the \Sigstar\ of the spaxel.
Given that \lowsf\ galaxies are predominantly characterized by fully-quenched regions at all radii, their \Sigsfr\ is notably lower compared to other populations, as shown in Figure \ref{fig_radial_phyprops}(a).
While the majority of \lowsf\ galaxies exhibit low \Sigsfr\ across their disks, one galaxy (7977-9101) displays high \Sigsfr\ in its central region. 
This is attributed to a relatively low D4000 value ($\sim$ 1.2, comparable to that of star-forming regions) in its center, which is identified as a \emph{composite} region based on the BPT analysis. 
The center of this galaxy is associated with an AGN \citep{Lin17}. Typically, such composite AGNs are commonly linked with young stellar populations and low D4000 values, indicative of a recent cessation of star-forming activities \citep[e.g.,][]{Wan08}. 
The timescale of quenching is beyond the scope of this paper, but we aim to delve into this question in future studies, focusing not only on the ALMaQUEST sample but also extending the analysis to include the entire MaNGA sample.

The proportion of retired regions (with D4000-based \Sigsfr), albeit at a low level, also increases from cSB to \lowsf\ galaxies (pink curve), though not as substantially as observed for fully-quenched regions. 
The occurrence of AGN-contaminated regions (also with D4000-based \Sigsfr) exhibits no discernible trend with galaxy populations and remains at a low level, below 20\%, across all populations (magenta curves).
Hence, the contribution of retired regions and AGN-contaminated regions to Figure \ref{fig_radial_phyprops}(a) is comparatively lower in contrast to star-forming and fully-quenched regions.

Taking all of these factors into account, we can deduce that the offsets of \lowsf\ from more star-forming galaxies are likely to be real \citep[e.g.,][]{Blu20b}, but we refrain from drawing strong conclusions regarding the exact profile and magnitude of this offset.
Also, some green-valley galaxies in the ALMaQUEST sample have high inclination (see Figure 2 in \citealt{Lin22}).
While the inclination and position angle of a galactic disk  are taken into account during the construction of the radial profile for individual galaxies, accurately deriving azimuthally-averaged radial profiles of \Sigsfr\ (as well as other physical properties) remains challenging for highly-inclined galaxies \citep{Iba19,Bar22}.

Turning to \Sigstar\ in Figure \ref{fig_radial_phyprops}(b), our galaxies show a strong and smooth dependence of \Sigstar\  on galactic radius. 
The median profiles of \Sigstar\ exhibit significant overlap among the different populations across all the radii explored.
This is somewhat expected, as our four populations share the similar range in global \Mstar\ as shown in Figure \ref{fig_sample}  \citep[][]{Gar17}.
Due to the uniform distribution of \Sigstar, any observed differences in star formation and gas properties among the galaxy populations (if present) are more likely to be genuine rather than a result of systematic dependence on stellar mass.

Figure \ref{fig_radial_phyprops}(c) displays the radial distribution of \Sigmol. 
For our galaxies, while there are no dramatic differences observed in the median profiles among the galaxy populations, there is significant individual variation in \Sigmol, ranging from one to two orders of magnitude from galaxy to galaxy. 
This variation can be attributed, at least in part, to the structured and non-smooth disk  of the ALMaQUEST galaxies and the clumpy nature of molecular gas \citep[][]{Ler13,Stu23}.
The dark blue and blue curves in Figure \ref{fig_radial_spxtype} represent the fractions of spaxels with CO detection and those without detection (upper limits), respectively.
For cSB, \highsf, and \midsf\ galaxies, the relative proportions of these two types of spaxels exhibit a considerable degree of similarity across the radial range explored.
However, it is noteworthy that \lowsf\ galaxies display notably lower \Sigmol\ values than the other galaxy populations for $R$ $<$ 1.0\Reff. 
This reduction in \Sigmol\ is attributed to the significant presence of CO non-detection in this region, as illustrated in Figure \ref{fig_radial_spxtype} (panel (d)). 
This is particularly true for the outer regions.
Beyond 1.0\Reff, the radial profile of \Sigmol\ flattens due to the increased prevalence ($>$ 50\%) of CO non-detection in this regime.
However, it is worth noting that in the inner region, specifically at $<$ 0.5\Reff, the detection of CO emission remains relatively robust in  \lowsf\ galaxies, even in the presence of low \Sigsfr.

To summarize, the largely overlapping radial profiles of \Sigmol\ (shaded lines in Figure \ref{fig_radial_phyprops}(c)) suggest that galaxies with the lowest levels of star formation activity (\lowsf) do not consistently display a significant deficit in molecular gas. 
However, since there is considerable variation in the \Sigsfr\ profiles (Figure \ref{fig_radial_phyprops}(a)), this implies that there must be a radial variations in \fgas\ and SFE, which will be further discussed in the next section.

Finally, while our primary focus is on the median profiles, it is important to note the substantial variation from galaxy to galaxy within a given category. 
This variation is particularly pronounced for \Sigsfr\ and \Sigmol, but less so for \Sigstar. 
The diverse morphologies of the galaxies in the ALMaQUEST sample contribute to these diverse profiles due to varied distributions of the clumpy star-forming regions and molecular clouds across their disks \citep[][]{Ler13,Pan22,Stu23}.
Our sample includes a broad spectrum of Hubble types, ranging from T-types of -1.6 to 6.5 (S0 to Scd), and features both barred and non-barred galaxies along with different spiral patterns\footnote{The T-type values for the ALMaQUEST galaxies are taken from the MaNGA Deep Learning Morphology Value Added Catalogue \citep[MDLM-VAC-DR17][]{Dom22}. 
This catalog has been developed through an automated classification system that employs supervised deep learning. 
The Hubble T-type classification primarily takes into account the ellipticity and the prominence of spiral arms, but it does not specifically account for the presence or absence of a galactic bar. 
The identification of a galactic bar and the further  verification of spiral arms are based on the updated Galaxy Zoo classification \citep{Wal22} and by our kinematic analysis in the forthcoming study by Lopez-Coba et al. (in preparation). 
Additionally, several galaxies display asymmetric disks, suggesting a history of past interactions \citep{Ell23}. 
Our current sample size is too limited to systematically explore the potential effects of morphology and environment on the radial profiles. 
However, such analysis can be conducted in a statistically significant manner using the full MaNGA sample.
This will be similar to what has been done with the CALIFA galaxies, albeit with a smaller sample size \citep{Gon16,San21}.
}.

\subsubsection{Radial Profiles of \fgas, SFE, and \Sigssfr}
\label{sec_result_rad_ratios}

In this section, we present the radial profiles of various physical properties derived from  \Sigsfr, \Sigmol, and \Sigstar. 
Specifically, we present the radial specific SFR (\Sigssfr) defined as \Sigsfr/\Sigstar.
We note that the symbol for \emph{surface density}, $\Sigma$, is not necessary in the expression for the radial (local) specific SFR as the ``per area'' component cancels out. However, in order to distinguish between the \emph{global} and \emph{local} sSFR in this paper, we retain the $\Sigma$ in the symbol. 
Additionally, we examine the molecular gas fraction (\fgas) defined as \Sigmol/\Sigstar\ and the star formation efficiency (SFE) defined as \Sigsfr/\Sigmol.
Figures \ref{fig_radial_ratios}(a)-(c) illustrate the radial profiles of \Sigssfr, \fgas, and SFE, respectively. 
The color scheme and line styles employed are the same as for those used in Figure \ref{fig_radial_phyprops}.
We will begin by examining the results obtained from the non-cSB sample, followed by an examination of the results from the cSB population.

As galaxies move from the star-forming main sequence to lower SFR regimes, the radial profiles of \Sigssfr\ exhibit a dependency with respect to global sSFR, displaying a decreasing trend in magnitude from \highsf\ to \lowsf\ (Figure \ref{fig_radial_ratios}(a)). 
This dependence agrees with previous studies conducted on  global scale  \citep[e.g.,][]{Sai17,Pio20}.
Furthermore, the decline in local \Sigssfr\ is observed consistently from the galactic center to 1.5~\Reff, indicating that star formation quenching is a phenomenon that occurs globally within galaxies \citep[see also][]{Cat17,Bel18,Blu20a,Blu20b}. 

The radial profiles of \highsf\ and \midsf\ galaxies show in Figure \ref{fig_radial_ratios}(a) exhibit remarkable similarity to that
 presented in \cite{Blu20b}, which were derived using the complete MaNGA DR15 sample, and with the fully-quenched regions considered.
The comparison validates that the \Sigssfr\ trends of the ALMaQUEST sample, despite its smaller size, is representative of that from larger sample of galaxies.
However, the profiles in Figure \ref{fig_radial_ratios}(a) are only \emph{qualitatively} consistent with the findings of \cite{Bel18} and \cite{Ell18} using full MaNGA sample and \cite{Gon16} using different IFU dataset  from the CALIFA survey.
While these studies did demonstrate a dependence of local \Sigssfr\ on global sSFR, the radial profiles of \Sigssfr\ exhibit some discrepancies.
This discrepancy arises because \cite{Bel18}, \cite{Ell18}, and \cite{Gon16} did not account for fully-quenched regions in their analyses. 
For instance, in \cite{Ell18}, the quiescent population display an increase in \Sigssfr\ with radius, as evident in their Figure 8, whereas in this work and \cite{Blu20b}, the profile remains relatively constant. 
The underlying reason for this contrast is that our work and \cite{Blu20b} take into consideration the presence of fully-quenched regions, whereas \cite{Ell18} (and other studies) primarily focused on the regions with H$\alpha$ (emission lines) detection.
In short, these studies demonstrate that the \emph{observed}  radial \Sigssfr\ is sensitive to  the choice of methodology, namely, between completeness and accuracy.

From \highsf\ to \midsf\ galaxies, the observed decrease in \Sigssfr\  with decreasing radius, as illustrated in Figure \ref{fig_radial_ratios}(a), also implies an inside-out quenching scenario, in which star formation ceases in the inner region of the galaxy more pronouncedly than the outer regions.
The inside-out quenching scenario is  in agreement with  previous studies  that utilized photometric colors, such as $NUV-r$ or UV-optical gradients, as indicators of star formation profiles \citep{Pan16,Liu16,Wan17}. 
Additionally, studies employing IFU (MaNGA, CALFA, and SAMI) with \Sigssfr\ based on emission lines analysis have yielded  consistent results \citep{Gon16,Bel18,Med18,San18,Ell18,Lin19a,San21,Vil23}.
Recently, analysis of the stellar-age gradients in late-type galaxies further  reinforces the notion of an inside-out cessation of star-forming activity in massive galaxies \citep{Bre20,Wan22,Lah23}.

Nevertheless, it is worth mentioning that \cite{Blu20b} demonstrated that  satellite galaxies with lower \Mstar\ (log(\Mstar/M$_{\sun}$) $<$ 10, i.e., lower than that of ALMaQUEST sample) undergoing quenching display rising star formation profiles,  indicating outside-in quenching, unlike high-mass central and satellites galaxies  which clearly quench inside-out. 
\cite{Lin19a}  also show the relative fraction of outside-in and inside-out quenching mode indeed strongly dependent on \Mstar, with high-\Mstar\ galaxies exhibiting a greater fraction of inside-out quenching compared to low-\Mstar\ ones.
Therefore, while the ALMaQUEST sample lacks low-\Mstar\ systems, it is important to note that the radial profiles of \Sigssfr\ shown in Figure \ref{fig_radial_ratios}(a) cannot be extrapolated to low-\Mstar\ population.

It is challenging to precisely quantify the extent of the decrease in \Sigssfr\ from \midsf\ to \lowsf\ due to inherent limitations in the  accuracy of \Sigssfr\ measurements, particularly for the \lowsf\ population. 
This limitation is also emphasized in the analysis conducted by previous observational studies \citep[e.g., ][]{Bel18,Ell18,Wan18,Blu20b}, despite their larger sample size. 
Nevertheless, it is apparent that \lowsf\ galaxies are experiencing star formation at a substantially lower rate across their entire disk, even after accounting for the normalization by stellar mass (i.e., \Sigssfr), indicating a global quenching scenario.

The radial profile of \fgas\ consistently shifts vertically downwards  decreasing global sSFR at almost all radii. 
Computed using \Sigstar\ and \Sigmol, \fgas\ demonstrates less sensitivity to completeness and accuracy issues compared to \Sigsfr-related quantities. 
\highsf\ and \midsf\ galaxies   show a mild increasing trend with radius within the $R$ $<$ 1.0\Reff\ regime. 
However, it is important to note again that caution should be exercised when interpreting the radial profile of \fgas\ beyond 1.0\Reff\ in all populations, owing to the decreased accuracy of the medians due to the increase in spaxel without CO detection, as illustrated in Figure \ref{fig_radial_spxtype}.
For \lowsf\ galaxies, we are not able to precisely quantifying their \fgas\ profile due to a significant number  of CO non-detections. 
Nonetheless, it is evident that \lowsf\ galaxies consistently exhibit lower gas content compared to the other populations across all radii considered.
A similar conclusion is reached by other analysis based on  CO analyses, as well as those using the dust-to-gas relation as a proxy for  molecular gas mass \citep{Sai17,San18,Lin19b,Col20,Bro20,Pio22,Lin22}.
Our finding is further corroborated by the most recent analysis of \fgas\ radial profiles for green valley galaxies using EDGE-CALIFA data by \cite{Vil23} (see their Figure 12), but note that this study utilizes a smaller sample size.

The radial distribution of SFE  also exhibits a clear ranking with global sSFR in Figure \ref{fig_radial_ratios}(c). 
The median log(SFE/yr$^{-1}$) for \highsf\ galaxies ranges from approximately $-$9.5 to $-$9.0, showing a rather flat profile  from the galactic center to the outer regions. 
The derived constant SFE for \highsf\ galaxies agrees with that of nearby large normal star-forming galaxies observed at kpc scales \citep[][]{Ler08,Big11,Mur19,Sun23}.
While the disk SFE of \midsf\ is lower than that of \highsf\ by only 0.2 dex, a central suppression of SFE is seen, as a result of a decease (increase)  in the  amount of star-forming (fully-quenched) regions (Figure \ref{fig_radial_ratios}).
A recent study by \cite{Vil23}, utilizing EDGE-CALIFA data, also observes a systematic increase of SFE with increasing radius in green valley galaxies; in contrast, the SFE in main sequence galaxies remains almost constant across their disks.
Although the  radial profile of SFE for \lowsf\ galaxies cannot be determined precisely due to the  uncertainty associated with \Sigsfr\ measurements, it is evident that the SFE of molecular gas in \lowsf\ galaxies is significantly suppressed by at least an order of magnitude compared to normal star-forming galaxies (\highsf).
A similar conclusion is also reached by other studies on both global and radial SFEs  \citep{Sai17,San18,Lin19b,Col20,Bro20,Pio22,Lin22}.

Mathematically, the decrease in  \fgas\ and SFE  can be attributed to an increase in the denominator (\Sigstar\ and \Sigmol, respectively) or a reduction in the numerator (\Sigmol\ and \Sigsfr, respectively).
Figure \ref{fig_radial_phyprops}(b) provides evidence that, statistically, radial \Sigstar\ remains relatively consistent across the sample.
Consequently, any variations in \fgas\ primarily arise from changes in \Sigmol\ (as also shown in Figure 9 of our previous paper in \citealt{Lin22}).
In terms of SFE, both \Sigmol\ and \Sigsfr\ exhibit lower values for \lowsf\ galaxies compared to \midsf\ and \highsf\ galaxies.
Therefore, the observed variations in SFE within our sample are a combined outcome of changes in both \Sigmol\ and \Sigsfr.

Moving to the cSB galaxies, it is evident that their median \Sigssfr\ magnitudes are higher than those of other populations at all radial bins (Figure \ref{fig_radial_ratios}(a)). 
Specifically, compared to galaxies located around the star-forming main sequence (i.e., \highsf\ galaxies in our sample), cSB galaxies exhibit an enhancement in \Sigssfr\ of approximately 0.5 dex in the innermost region and a consistently smaller enhancement of approximately 0.3 dex in the $R$ $>$ 0.75 \Reff\ regime. 
We note again that the significant central enhancement in \Sigssfr\ is a result of sample selection (see \S\ref{sec_sample_classification}).

Figure \ref{fig_radial_ratios}(b) and (c) shows   enhancement in both \fgas\ and SFE for cSB galaxies, the largest enhancements are seen in the inner regions ($R$ $<$ 0.5 $R_\mathrm{e}$), which corresponds to the region where \Sigssfr\ is most enhanced. 
This similarity suggests a potential link between the enhanced \Sigssfr\ and  SFE and/or \fgas\ in cSB galaxies.
We refer readers to \citet{Ell20a} for a more detailed analysis and the implications on the link between  \Sigssfr,  SFE and \fgas\ in cSB.

Finally, it is important to note that star formation and molecular gas can extend beyond the galactocentric radius examined in this study (1.5$R_\mathrm{e}$), as highlighted in previous works \citep[e.g.,][]{Wat16,Bac20,Kod22}. 
IFU observations conducted by MaNGA and CALIFA have indicated that \Sigssfr\ beyond the radius of $>$ 1.5$R_\mathrm{e}$ is typically lower than that within 1.5$R_\mathrm{e}$ \citep[e.g.,][]{Gon16,Bel18,San21}.
Hence,  our results are not expected to be strongly influenced by star formation in the outskirts.
Furthermore,  the comparison of radial profiles between different galaxy populations is valid in this study due to the relatively narrow range of global \Mstar\ within the ALMaQUEST sample. 
Previous studies have shown that the radial profiles of star formation and (molecular) gas properties, including their shapes and magnitudes, depend on the \Mstar\ of galaxies \citep{Gar17,Bel18,San21,Bar22}. 
Therefore, such comparisons between galaxy populations (sub-samples) should be approached with caution when dealing with a wide range of mass values.

\subsection{Relative Importance of \fgas\ and SFE in Quenching}
\label{sec_relative_importance}

Many previous studies have assessed the relative role of \fgas\ and SFE for \emph{boosting} star formation.
For example, on global scale, there exists a correlation between global SFE and distance above the main sequence \citep[e.g.,][]{Sar14,Gen15,Sai17,Tac18}.
Conversely, other studies have shown that increased \fgas\ plays a pivotal role in regulating star formation \citep[e.g.,][]{Sco16,Lee17}.
\cite{Pan18b} also found a correlation between the level of merger-triggered starburst and global \fgas.
To avoid the potential dilution of key physical processes that may occur when averaging properties across entire galaxies, in our previous work, we quantify the relative role of \fgas\ and SFE for the cSB sample in the ALMaQUEST survey on a spatially resolved manner \citep{Ell20a}. 
Our findings suggest that an elevated SFE is the primary driver of central starburst activity.
We also investigate the mechanisms responsible for merger-triggered star formation using the ALMaQUEST merger sample \citep{Tho22}. 
The analysis indicates that both \fgas\ and SFE can drive merger-triggered star formation.
In this work, our focus shifts to assessing the relative significance of \fgas\ and SFE in the context of suppressing, rather than boosting, star formation.

To gain a deeper understanding of the factors influencing the radial profiles of \Sigssfr\ in quenching galaxies, we compare the extent of quenching with the changes in \fgas\ and SFE relative to the normal star-forming galaxies (\highsf) within our sample.
More specifically, for each individual galaxy classified as \midsf\ or \lowsf, we first calculate the deviation of their \Sigssfr\ (i.e., $\Delta$\Sigssfr) from the median \Sigssfr\ of all \highsf\ galaxies (represented by the thick blue curve in Figure \ref{fig_radial_ratios}(a)) at each radial bin.
Similarly, we compute the radial $\Delta$\fgas\ and  $\Delta$SFE using the same approach.
Then the relative significance of SFE and \fgas\ can be quantified by comparing $\Delta$\fgas\ and $\Delta$SFE, specifically the value of the expression $\Delta$\fgas$-$$\Delta$SFE. 
This expression allows for a ranking of the two drivers \citep[e.g.,][]{Ell20a,Bro20,Mor21,Tho22,Gar23}, revealing which one exerts a more pronounced influence on quenching star formation at different galactocentric radii. 
A positive value for this expression indicates that the deviation of \Sigssfr\ from that of the main-sequence galaxies is \emph{primarily, but not necessarily exclusively}, driven by an increase in SFE. 
Conversely, a negative value signifies a scenario \emph{primarily} influenced by \fgas.
Note that regardless of which quenching driver plays the dominant role, we do not deny the significance of the other one.
The uncertainty of $\Delta$\fgas$-$$\Delta$SFE is  determined through error propagation from  the uncertainties of the raw curves.

\begin{figure*}
	\centering
	\includegraphics[scale=0.65]{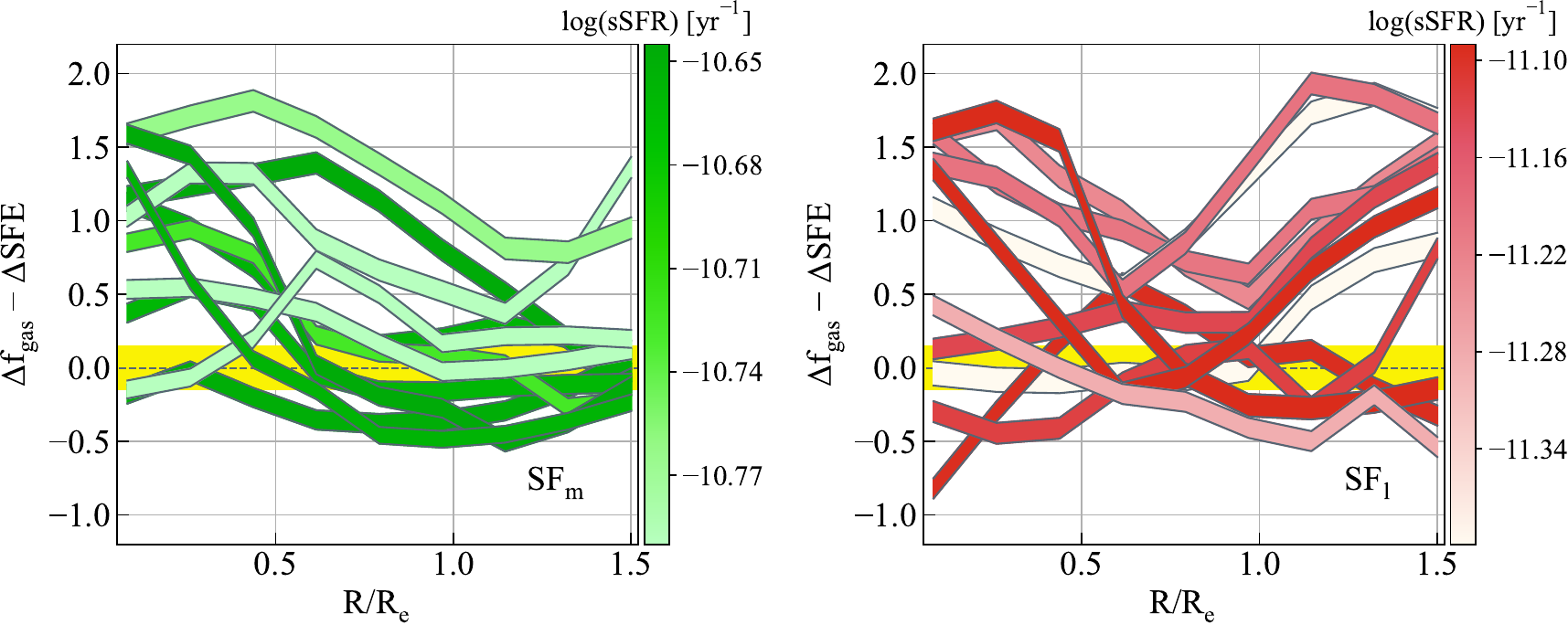}	
	\caption{Radial profile of 	$\Delta$\fgas$-$$\Delta$SFE for \midsf\ (left) and \lowsf\ (right) galaxies. A positive value for this expression indicates that the deviation of \Sigssfr\ from that of the main-sequence galaxies is primarily driven by an increase in SFE. A negative value suggests a scenario primarily influenced by \fgas. The darkness of the curve color increases with global sSFR values. The yellow shaded region represents the regime ($-$0.15 $<$ $\Delta$\fgas$-$$\Delta$SFE $<$ 0.15) where \Sigssfr\ is determined by the combined effect of both \fgas\ and SFE. 
	}  
	\label{fig_rad_driver}
\end{figure*}

Figure \ref{fig_rad_driver} displays the radial profiles of $\Delta$\fgas$-$$\Delta$SFE for \midsf\ galaxies on the left and \lowsf\ galaxies on the right. 
The galaxies are color-coded based on their global sSFR, which is  related to their distance from the main sequence. 
The darkness of the curve color increases with global sSFR values.
The yellow horizontal shaded region represents the regime ($-$0.15 $<$ $\Delta$\fgas$-$$\Delta$SFE $<$ 0.15) where \Sigssfr\ is determined by the combined effect of both \fgas\ and SFE. 
The limit of $\pm$0.15 is approximately twice the typical uncertainty of the $\Delta$\fgas$-$$\Delta$SFE profiles.
The choice of this limit should be considered as a conservative choice to  avoid the risk of over-interpretation.
While the majority of galaxies in both the \midsf\ and \lowsf\ categories tend to be skewed towards being primarily driven by SFE, there exists a diverse configurations in terms of the dominant factors influencing star formation.

\begin{figure*}
	\centering
	\includegraphics[scale=0.38]{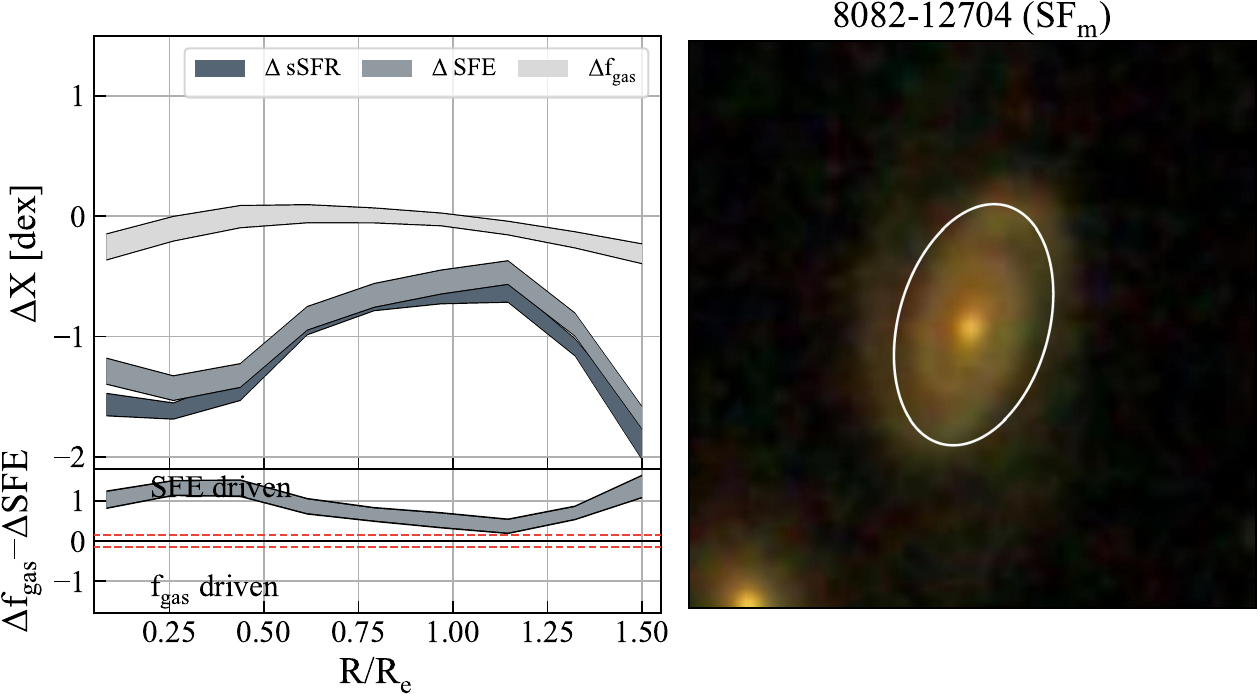}
	\includegraphics[scale=0.38]{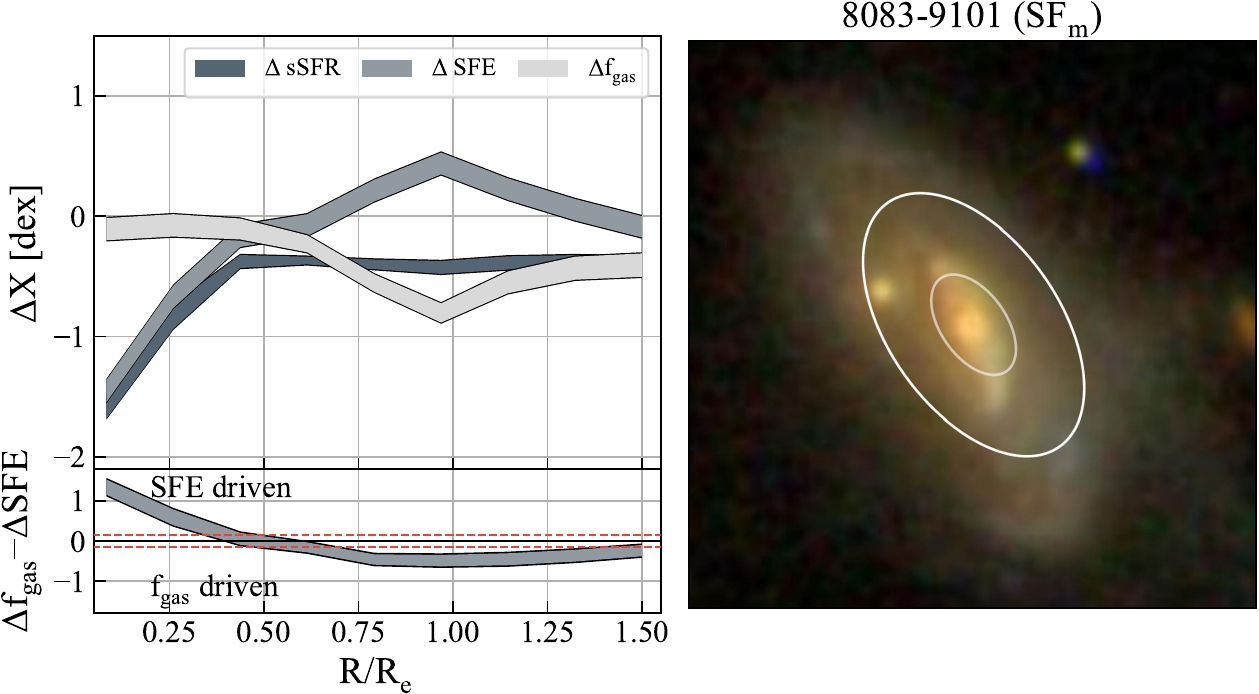}
	\includegraphics[scale=0.38]{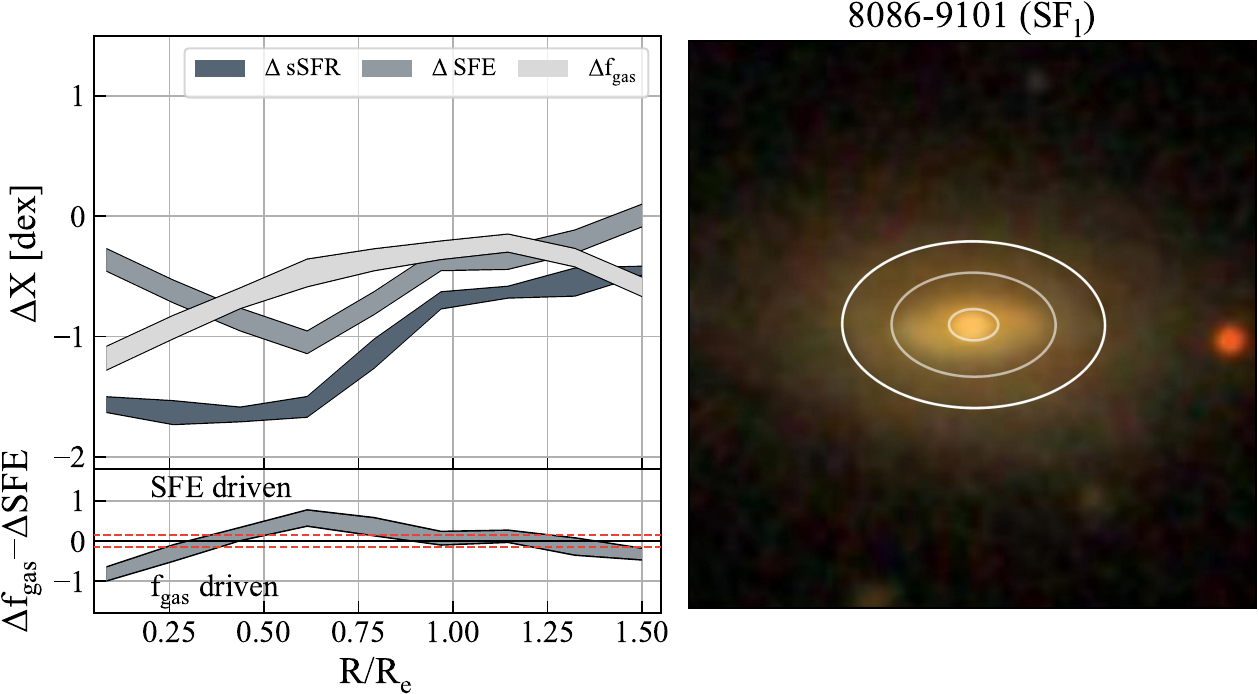}
	\includegraphics[scale=0.38]{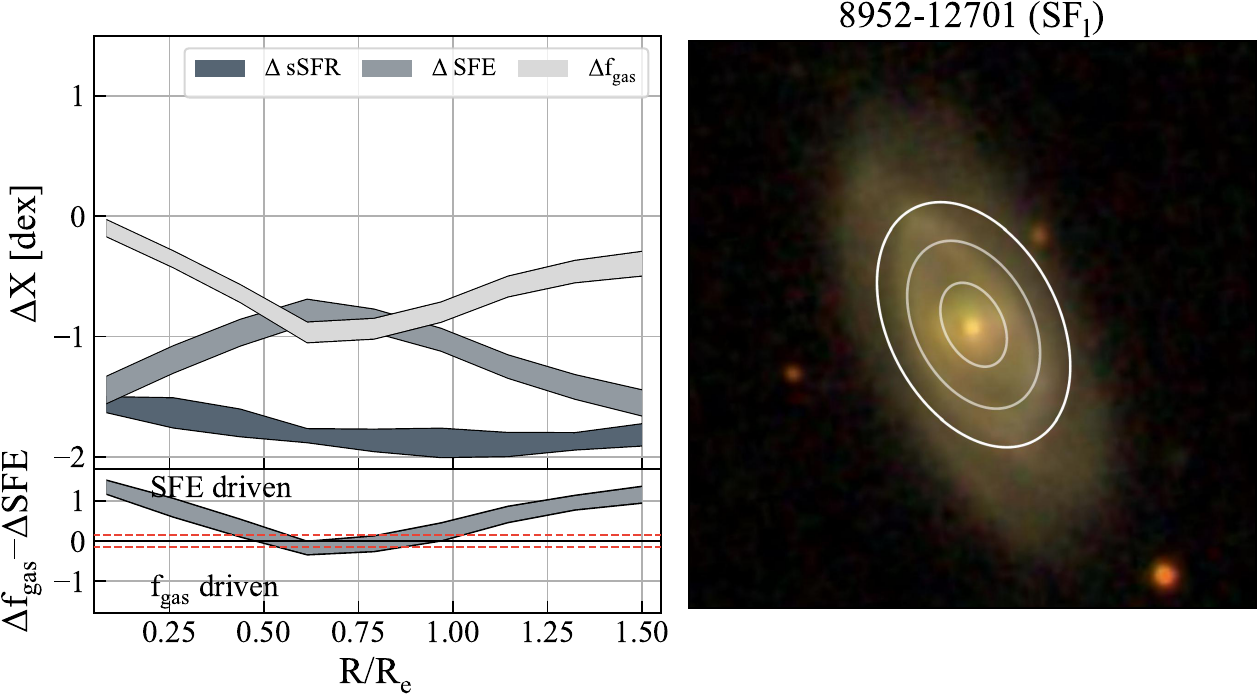}
	\caption{Examples of radial $\Delta$ properties for four galaxies exhibit diverse configurations of dominant drivers within their disks. In each galaxy, the upper-left corner displays gray curves, with increasing darkness representing $\Delta$\fgas, $\Delta$SFE, and $\Delta$\Sigssfr, respectively. The lower-left corner presents the radial $\Delta$\fgas$-\Delta$SFE profile. The red dashed lines mark the regime ($-$0.15 $<$ $\Delta$\fgas$-\Delta$SFE $<$ 0.15) where \Sigssfr\ is determined by the combined effect of both \fgas\ and SFE. The right panel shows the SDSS $gri$ composite image of the galaxy, with the outermost ellipse indicating the radius of 1.5\Reff. For galaxies exhibiting multiple dominant drivers at different radii (upper-right, lower-left, and lower-right galaxies), the inner ellipse(s) indicate the radii where the dominant driver changes.
	}  
	\label{fig_deltaX_example}
\end{figure*}

Figure \ref{fig_deltaX_example} provides four illustrative examples. 
In each sub-figure, the upper-left panel displays gray curves, with increasing darkness representing $\Delta$\fgas, $\Delta$SFE, and $\Delta$\Sigssfr, respectively. 
Meanwhile, the lower-left panel presents the radial $\Delta$\fgas$-$$\Delta$SFE profile. 
The right panel exhibits the SDSS $gri$ composite image of the galaxy, with the outermost ellipse highlighting the radius of 1.5\Reff.
The upper-left panel of Figure \ref{fig_deltaX_example} exemplifies a case (8082-12704) where the radial $\Delta$\Sigssfr\ is primarily influenced by SFE across the entire radial range. 
Conversely, other panels show galaxies that demonstrate signs of multiple mechanisms influencing their star formation. 
Specifically, in the case of galaxy 8083-9101 (upper-right), SFE dominates in the inner regions, while \fgas\ takes over in the outer regions, with a  transition occurring at approximately 0.6 \Reff.
The lower two panels show  galaxies (8086-9101 and 8952-12701) exhibiting a more complex pattern. 
Star formation in the inner regions (within 0.3~\Reff) of 8086-9101 is primarily driven by \fgas, while in the intermediate radial range (0.3 -- 1.0 \Reff), SFE takes the lead. 
In the outermost regime, both \fgas\ and SFE seem to play a role in shaping star formation.
For galaxy 8952-12701, star formation in the inner and outermost regions is primarily governed by SFE. 
However, in the intermediate radial range (0.5 – 1.0 \Reff), both \fgas\ and SFE work together to regulate star formation.

Figure \ref{fig_rad_driver} and \ref{fig_deltaX_example}   provides compelling evidence that star formation in quenching galaxies is governed by a range of diverse modes, SFE-driven, \fgas-driven, and by the combination of the two; such complexity  is also observed in galaxies experiencing enhanced star formation \citep{Ell20b,Tho22}.
Given the small size of our sample in this work, it is challenging to pinpoint the exact underlying origin of the multiple-mode drivers, such as different underlying mechanisms being responsible for star formation quenching in different galactocentric regimes. 
Further studies with larger sample sizes and increased statistical significance are necessary to provide more robust constraints on the prevalence  and the nature of such multi-mode driver. 
Moreover, employing different \Sigsfr\ tracers that are insensitive to contamination from AGNs and older stellar populations, or are unaffected by dust attenuation, will contribute to confirming our findings based on the H$\alpha$ and D4000 star formation measurements.

\subsubsection{Relative Importance of \fgas\ and SFE in Individual Galaxies}
\label{sec_relative_importance_individual}
In this section, we aim to discuss the spatial characteristics of individual galaxies. 
Using the radial profiles of $\Delta$\fgas-$\Delta$SFE  in Figure \ref{fig_rad_driver}, we undertake a visual classification of galaxies into two distinct groups based on the number of dominant factors influencing star formation within their  galactic disks. 
These groups are characterized as either ``single mode''  or ``multiple driver''.
For example, the galaxy shown in the upper-left panel of Figure \ref{fig_deltaX_example} would be classified as a ``single mode'' system, driven primarily by SFE. 
In contrast, other galaxies in Figure \ref{fig_deltaX_example}  would fall into the ``multiple driver'' category, each displaying a complex combination of factors influencing their star formation.

Using this classification approach, we identify 10 galaxies (4 \midsf\ and 6 \lowsf; 1 \fgas-driven and 9 SFE-driven) categorized as being driven by a single mode and 12 galaxies (7 \midsf\ and 5 \lowsf) classified as being influenced by multiple modes. 
There is no significant difference in the number of galaxies between these two modes, and there is no apparent trend regarding the dominant mode of quenching for \midsf\ and \lowsf\ galaxies.
By combining the previous ALMaQUEST results on enhanced star formation \citep{Ell20b,Tho22}, we conclude that \emph{both quenching and star formation boosting is both individualized (from galaxy-to-galaxy) and can be complex even with a given galaxy}.

\subsubsection{Relative Importance of \fgas\ and SFE at Different Radial Range}
\label{sec_relative_importance_all}
\begin{figure}
	\centering
	\includegraphics[scale=0.6]{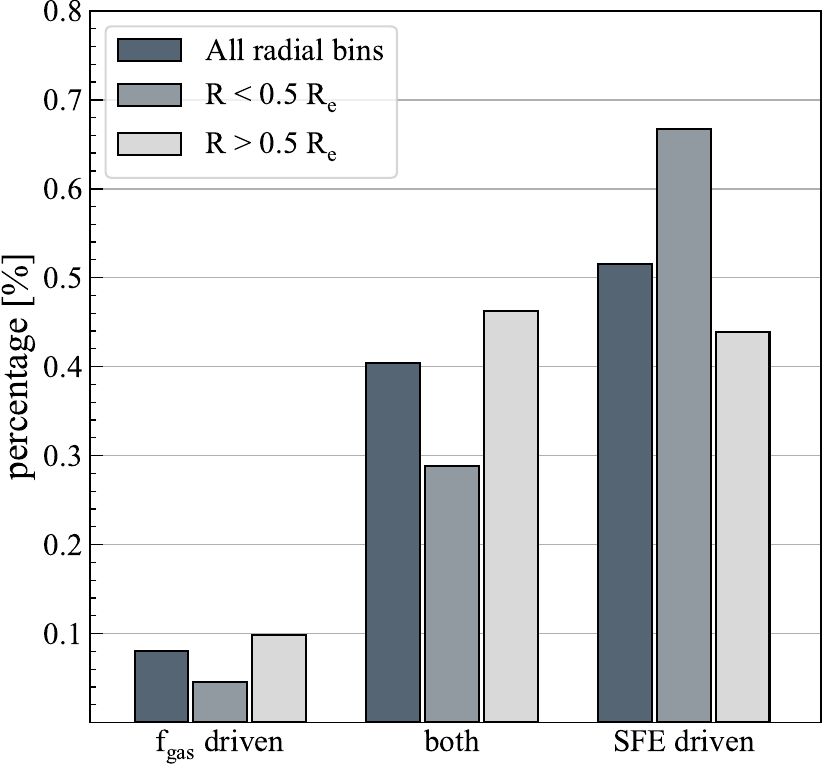}
	\caption{Distribution of different drivers across 198 azimuthal bins in the 22 quenching galaxies. When the value of $\Delta$\fgas$-$$\Delta$SFE is exactly zero, or when the associated uncertainty encompasses $\pm$0.15, the region is considered to be equally influenced by both drivers. The dark-gray histograms represent the distribution for all radial bins, while the gray and light-gray histograms represent the distributions for the central ($R$ $<$ 0.5\Reff) and  disk ($R$ $>$ 0.5\Reff) regions, respectively.
	}  
	\label{fig_driver_hists}
\end{figure}

In the previous section, we discussed the individual behavior of the 22 galaxies below the main sequence (\midsf\ and \highsf), in this section, we will  answer the question of whether there is a preference for a specific quenching driver (\fgas, SFE, or a combination of both) at different galactocentric radii.
Figure \ref{fig_driver_hists} illustrates the distribution of the different drivers across 198 azimuthal bins in the 22 quenching galaxies. 
When the value of $\Delta$\fgas$-$$\Delta$SFE is precisely zero or when the associated uncertainty encompasses $\pm$0.15, the region is considered to be equally influenced by both drivers.
First, we will examine all the radial bins as a whole. 
Moreover, since the central region ($R$ $<$ 0.5\Reff) exhibits the most pronounced quenching as galaxies begin to deviate downwards from the main sequence (Figure \ref{fig_radial_ratios}(a); see also \citealt{Ell18}, \citealt{Bel18}, and \citealt{Blu20b}), it merits distinct discussion and is separated from the remainder of the disk for discussion purposes.
The definition of the central region as $R$ $<$ 0.5\Reff, typically dominated by galactic bulges \citep{Kal21,Kal22},  is also consistent with the analysis taken in previous ALMaQUEST papers \citep{Ell20a,Lin22}.
Correspondingly, the disk region is defined as the area beyond  0.5\Reff.

Considering all the radial bins collectively (as represented by the dark gray bars in Figure \ref{fig_driver_hists}), SFE emerges as the dominant factor, accounting for approximately 51\% in determining the \Sigssfr. 
Following closely behind is a mode where \fgas\ and SFE contribute equally, constituting approximately 40\%, while cases primarily driven by \fgas\ account for only 9\%.
In the  central region (consisting 66 radial bins for $R$ $<$ 0.5 $R_\mathrm{e}$; gray bars in Figure \ref{fig_driver_hists}), the trend aligns with the overall dataset, with SFE driving 66\% of radial regions, the combination of \fgas\ and SFE playing a significant role in 29\% of the regions, and \fgas\ being the primary driver in 5\% of cases.
Our findings regarding the   central region are consistent with the results of an integrated analysis conducted by \cite{Col20}. 
Their study, based on EDGE-CARMA and APEX observations for approximately 470 galactic centers, similarly suggests that as galaxies depart from the main sequence, it is the changes in SFE that primarily push a galaxy (more specifically,  their quenching center) further into the passive regime.
It is important to note that the term ``center'' as used in \cite{Col20} corresponds to $\sim$~1.0\Reff, defined by the APEX beam at 230 GHz. 
This is  larger than the definition of ``center'' used in our work. This distinction in spatial definitions should be taken into account when comparing results across different studies.
We have evaluated our results using an alternative definition of the center as $R$~$<$ 1.0\Reff. 
While this adjustment alters the fraction of individual categories in Figure \ref{fig_driver_hists}, the relative differences among the  various categories remain consistent.

In the disk region (consisting of 132 radial bins for $R$ $>$ 0.5\Reff; light gray bars), a different pattern emerges. 
Here, \fgas\ and SFE are equally important in driving star formation in 47\% of the radial bins, becoming the most prevalent mode in this radial regime. 
At the same time, SFE takes the lead as the dominant factor in 43\% of these radial bins, with \fgas\ remaining the least common driver, accounting for only 10\% of the regions in the disk region.
Several prior studies, relying on integrated measurements of entire galaxies, have demonstrated the significance of both decreasing \fgas\ and SFE in determining the distance from the star-forming main sequence \citep[e.g.,][]{Sai17,Pio20}. 
While these studies cannot distinguish between the central and disk regions of galaxies, their findings are presumably heavily influenced by the events transpiring in the disk regions. 
This is due to the extensive size of these regions, which therefore contributes significantly to the overall average quantity.
Consequently, our findings that the combination of \fgas\ and SFE is the most prevalent mode in the disk regions can offer an explanation for the findings of previous integrated studies.

In our analysis, we find that star formation in the radial regions considered is primarily governed either by SFE or by the combined influence of \fgas\ and SFE. 
The implication is consistent to that reported in previous spatially-resolved analysis \citep{Lin17,Lin22,Bro20}, namely that, statistically,  both \fgas\ and SFE contribute to the sSFR suppression in  central and disk regions, but the relative importance between the \fgas\ and SFE in lowering the \Sigssfr\  vary from galaxy to galaxy.
It is essential to note that while \fgas\ may not emerge as the statistically dominant driver of star formation in any of the regions we have examined, this result does not negate the fact that \fgas\ is indeed reduced in quenching galaxies. 
Instead, what becomes evident is that the magnitude of the reduction in \fgas\ alone is not substantial enough to account for the degree of the suppression observed in star formation.

\section{Discussion}
\label{sec_discussion}
The analysis of the radial profiles in this study, as shown in Figure \ref{fig_radial_ratios}, provides a direct visualization of the significance of both \fgas\ and SFE  in shaping the star formation behavior in galaxies undergoing quenching. 
The clear ranking of radial \fgas\ and SFE values in relation to the distance from the global star-forming main sequence, observed across all radial ranges up to 1.5$R_\mathrm{e}$, suggests a gradual suppression of both \fgas\ and SFE as galaxies moving downwards from the main sequence. 
This finding is consistent with our previous work within the ALMaQUEST series \citep{Lin17,Lin22}, as well as other spatially-resolved studies \citep[e.g.,][]{Bro20,San21,Vil23}.

\subsection{Quenching Mechanisms}
\label{sec_quech_mechanisms}

The diverse configuration  of dominant quenching drivers, observed both within individual galaxies and across all 22 quenching galaxies, points to a complex set of quenching mechanisms operating within our galaxy sample.
Several mechanisms have been proposed to operate in the central regions of galaxies. 
For instance, feedback from star formation \citep[e.g.,][]{Mur05,Hop14} and AGNs  \citep[e.g.,][]{Sij07,Hla12,Sch15,Wei18} can disrupt and disperse star-forming clouds, reducing the \fgas\ \citep[e.g.,][]{Ell21c,Gar21} and consequently lowering \Sigssfr.
Conversely, gas in  galaxies can also be  supported thermally by AGN heating or by  the non-thermal pressure from the magnetic field, cosmic rays and turbulence to counteract gravitational collapse \citep{Tab18}.
These effects can lower the SFE while not necessarily reducing \fgas.
Figure \ref{fig_radial_spxtype} shows that AGN-contaminated spaxels exist in all populations, but there does not appear to be a clear relationship between the fraction of AGN-contaminated spaxels and the galaxy's deviation from the main sequence within our sample. 
Nonetheless,  there might be a significant time-lag between the initiation of star formation quenching and the emergence of observable AGN signatures \citep{Sch09}.

Beyond the central regions of galaxies, in the context of the morphological quenching scenario \citep{Mar09}, the formation of a bulge acts to stabilize the gaseous disk of the galaxy, preventing its collapse (i.e., reduction of SFE), without necessarily requiring a mechanism to remove the gas within it \citep{Col18,Men19}.
On the other hand,  molecular gas within the galactic disk can be removed when the ram pressure of the ISM exceeds the gravitational force provided by the galaxy's disk \citep{Gun72}. 
This removal leads to a reduction in \Sigssfr\ due to the loss of molecular gas content (\fgas) \citep{Bos14a,Bos16,Sim18,Owe19,Bro23}.

The presence of a stellar bar in galaxies also introduces a complex interplay in star formation quenching \citep{Mas10,Fra20}. 
The bar-induced torque drives gas inflows toward the galactic center, enhancing the gas content and star formation in the central region (this phenomenon in fact contrasts with the \emph{quenching} process we have observed for both molecular gas and star formation here) while depleting the bar region of the gas (\fgas) necessary for star formation \citep{Kun07,Ell11,Zho15,Spi17,Kho18,Cho19,Geo21}.
Conversely, the bar can also induce shocks and shear, which increase the turbulence of the gas in the bar region; this, in turn, stabilizes the  gas against gravitational collapse, leading to the suppression of disk star formation by lowering the SFE \citep{Rey98,Ver07,Hay16}.
A subset of the ALMaQUEST galaxies has been visually identified as barred galaxies through the Galaxy Zoo project; there is also an increasing possibility (based on Galaxy Zoo) of a galaxy being a barred one from \highsf, to \midsf, and to \lowsf\ galaxies.
In our forthcoming papers (Hogarth et al. submitted, Lopez-Coba et al. in preparation),  we will investigate the properties of these bars in relation to molecular gas and star formation characteristics to gain a deeper understanding of their role in star formation quenching.

Within the ALMaQUEST sample, a few galaxies exhibit asymmetric optical disks, indicating past or ongoing interactions with other galaxies. Previous studies have demonstrated that galaxy interactions have a significant impact on star formation and cold gas properties \citep{Pan18b,Tho22,Gar23}.
However, the question of whether the merging process of galaxies can lead to the quenching of star formation remains a topic of debate \citep[e.g.,][]{Gab10,Wei17,Rod19,Dav22,Ell22,Qua23}. 
Nevertheless, a larger sample is certainly needed to distinguish between internal processes, such as bulge growth and bar formation, and external processes like galaxy interactions and mergers that regulate star formation and gas properties in galaxies.

While identifying the precise quenching mechanisms and their specific contributions to the observed \Sigssfr, \fgas, and SFE is beyond the scope of this paper, it is plausible that distinct quenching mechanisms  play a role in suppressing star formation across different radial regimes. 
These mechanisms could exert varying impacts on molecular gas content, resulting in the observed diversity in dominant quenching drivers (\fgas, SFE, or both) within our sample.
Furthermore, the underlying quenching mechanisms may operate over varying time scales.
There is a growing consensus that quenching is not necessarily  a rapid process \citep{Ell22} but instead unfolds gradually over a period of  1 -- 4 Gyrs \citep{Hah17, Fol18, Wri19, San19}. 
The exact timescale of quenching depends on various factors, including galaxy stellar mass (\Mstar), whether the galaxy is central or a satellite, and the particular underlying quenching mechanisms at play.

\subsection{Comparison with Simulations}
\label{sec_simulations}

Several recent studies have explored  \Sigssfr\ in simulated green-valley galaxies using galaxy simulations such as FIRE, Illustris, IllustrisTNG, EAGLE, and SIMBA \citep[e.g.,][]{Orr17,Sta19,App20,Nel21,Cor23a,Cor23b,Mcd23}.
Overall, most of the simulated galaxies below the main sequence exhibit lower \Sigssfr\ at nearly every radius compared to star-forming galaxies with similar \Mstar. 
The agreement between our observed radial \Sigssfr\ profiles and the trend observed in simulations strengthens the evidence for the global quenching phenomenon.
However, we also note that some simulations find centrally
enhanced sSFR, rather than the suppressed sSFR found in other simulations (and observations) (FIRE of \citealt{Orr17} and Illustris and EAGLE of \citealt{Sta19}).

While inside-out quenching mode is observed commonly in observations for high-mass galaxies (this work; \citealt{Bel18}; \citealt{Bro20}, ; \citealt{Blu20b}; \citealt{Nel21}; \citealt{San21}; \citealt{Vil23}), both inside-out and outside-in modes are predicted by simulations for galaxies with similar mass to our sample.
The outside-in mode is concluded to be as a result of a  strong
truncation in the sSFR profiles at $>$ 1\Reff\ for some simulated green-valley galaxies (Illustris and EAGLE of \citealt{Sta19} and SIMBA of \citealt{App20}).
The sharp cut-off is present in all AGN feedback variants; therefore, this may be originated from the adopted star formation (rather than AGN) prescriptions \citep{App20}.
Furthermore, it is important to note that there is no detection limit for \Sigsfr\ in simulations, unlike in observations. 
The approach of dealing with the observed fully-quenched regions (i.e., assigning an extremely low but not zero \Sigsfr) would naturally result in simulated galaxies experiencing quenching to a greater degree than our observational galaxies.

Our most-quenched green-valley galaxies (\lowsf) consistently demonstrate lower cold gas content (\Sigmol\ and \fgas) compared to other population groups,  consistent with findings from simulations  \citep{App20}. 
Nevertheless, simulations show a rapid decline in \Sigmol\ and \fgas\ at radial regimes below approximately $<$ 0.6\Reff\ for both main-sequence and green-valley galaxies, with the drop being more pronounced for green-valley galaxies.
However, our observed data do not exhibit this pronounced decrease in central \Sigmol\ and \fgas\ (Figure \ref{fig_radial_phyprops}(c) and \ref{fig_radial_ratios}(b)). 
This could be due to the overestimate of the feedback/energy injection of the simulated AGNs and their influence in the quenching \citep{Cor23a,Cor23b}.
On the other hand, the discrepancy of the central molecular gas  between simulated and observed galaxies could potentially be attributed to the assigned CO upper limits for the central quenched regions, which result in an overestimation of \Sigmol\ and, consequently, \fgas\ in observations. 
Nevertheless,  the accuracy of \Sigmol\ is presumed to be high, as the majority of spaxels in the central regions, even within the most-quenched population (Figure \ref{fig_radial_spxtype}), are detectable in CO.
Simulations have demonstrated that the gas content, including \ion{H}{1}, H$_{2}$, and CO,  of simulated main-sequence and green-valley galaxies is sensitive to feedback processes from both star formation and AGN \citep[e.g.,][]{Lag15,Cra17,Die19,Dav20,Ma22}.
Further investigation is required to understand the discrepancy between observations and simulations, but it is evident that  cold gas content and its distribution within a galaxy provide strong constraints on feedback processes and galaxy evolution models.

\section{Summary} 
\label{sec_summary}

We utilize  galaxies obtained from the ALMaQUEST survey \citep{Lin20} to investigate the variations in radial profiles of star formation and molecular gas properties. 
Our aim is to understand how these variations relate to star formation quenching. 
For this purpose, the 34 non-starburst galaxies in the ALMaQUEST main sample are categorized into three groups, including the star-forming main sequence (\highsf), galaxies  at the lower end of the main sequence scatter (\midsf), and galaxies residing significantly below the main sequence (\lowsf).
Additionally, we also explore whether there exists a preference for a particular quenching driver, such as gas fraction (\fgas) or star formation efficiency (SFE), at different galactocentric radii.
The 12 starburst galaxies in the ALMaQUEST survey were specifically selected to exhibit centrally peaked starbursts \citep{Ell20a}. 
These starburst galaxies are presented as a distinct category and are included in the relevant sections and figures for the purpose of comparison.

Our findings are as follows:
\begin{itemize}
	
	\item The median radial profiles of local stellar mass (\Sigstar) and molecular gas mass (\Sigmol) of each subset, derived by computing the median values from individual profiles, show no significant variation among the three populations. However, the radial profiles of local SFR (\Sigsfr) exhibit a  trend across the examined radial range, with a consistent increase in magnitude observed from the \lowsf\ galaxies to the \midsf, and finally to the \highsf\ galaxies at almost all radii (Section \ref{sec_radial_raw} and Figure \ref{fig_radial_phyprops}).

	\item Subsequently, we investigate the radial profiles of various physical properties derived from \Sigsfr, \Sigmol, and \Sigstar. Specifically, we analyze the radial distribution of the specific SFR (\Sigssfr\ $=$ \Sigsfr/\Sigstar), molecular gas fraction (\fgas\ $=$ \Sigmol/\Sigstar), and star formation efficiency (SFE $=$ \Sigsfr/\Sigmol). Our findings reveal a significant correlation in both magnitude and radial profile between these properties  and the transition of galaxies from the star-forming main sequence to lower SFR regimes. The radial profiles of \Sigssfr, \fgas, and SFE consistently exhibit a gradual decrease in magnitude from the \highsf\ group to the \lowsf\ group across almost all radii. These observed trends are not a result of any inherent dependence of gas and star formation properties on global \Mstar\ or local \Sigstar, as the selected galaxies have a relatively narrow range of \Mstar\ (10.2 -- 11.6 M$_{\sun}$ in log scale) and rather uniform  \Sigstar\ profiles. Our findings also support an inside-out quenching scenario for these relatively massive galaxies.  (Section \ref{sec_result_rad_ratios} and Figure \ref{fig_radial_ratios}).
	
	\item We quantify the relative importance of \fgas\ and SFE at different galactic radii for the galaxies undergoing quenching (\midsf\ and \lowsf). This is done by comparing the radial profile of   \fgas\ and SFE of individual quenching galaxies with the median profiles derived for the main sequence galaxies.  We found star formation in quenching galaxies is governed by a range of diverse modes, \fgas-driven, SFE-driven, or equally-driven (combination of \fgas\ and SFE). In roughly half of our quenching galaxies, star formation is predominantly  regulated by modes that operate across the entire galaxy -- whether it is \fgas, SFE, or a combination of both. 
	In contrast, the other half exhibits notable complexity, with different drivers influencing star formation at various radial ranges within these galaxies (Section \ref{sec_relative_importance_individual}, Figure \ref{fig_rad_driver} and \ref{fig_deltaX_example}). 
	
	\item When analyzing all galaxies together, statistical analysis reveals that the decline in star formation within the  central regions ($R$~$>$~0.5~\Reff) of galaxies is primarily attributed to a reduction in SFE. 
	Conversely, in the  disk regions ($R$~$<$~0.5~\Reff), both \fgas\ and SFE contribute to the suppression of \Sigssfr\ (Section \ref{sec_relative_importance_all} and Figure \ref{fig_driver_hists}).
	
	\item The diverse configuration of dominant quenching drivers, observed both within individual galaxies and across all 22 quenching galaxies, suggests that distinct quenching mechanisms may be responsible for suppressing star formation across different radial regimes. These mechanisms, likely operating over varying timescales, could have varying effects on molecular gas content, leading to the observed diversity in the dominant quenching drivers, including \fgas, SFE, or a combination of both, within our sample (Section \ref{sec_quech_mechanisms}).
	
	\item We compare the radial profiles of \Sigssfr\ and molecular gas properties observed in our study with those derived from galaxy simulations. While inside-out quenching is frequently observed in high-mass galaxies in our sample and other observations, simulations predict both inside-out and outside-in quenching modes for galaxies with similar masses to the ALMaQUEST sample. 
	Additionally, simulations indicate a  decrease in \Sigmol\ and \fgas\ at radial distances below 0.6\Reff\ for both main-sequence and green-valley galaxies, a trend not observed in the current observational data.
	These discrepancies is likely attributed to factors on both the simulation and observation sides, including the adopted AGN and star formation prescriptions, as well as the treatments of observational limitations and biases (Section \ref{sec_simulations}).
	
\end{itemize}

In summary, our results offer valuable insights into the evolution of radial star formation and molecular gas distribution as galaxies moving downwards from the star-forming main sequence.
 To validate our findings, a larger sample size is essential, enabling a more comprehensive analysis and the differentiation of various quenching mechanisms and their impact on star formation and molecular gas properties. 
 In a companion paper (\citealt{Lin24} ApJ in press), we further investigate the dense gas content traced by HCN and HCO$^{+}$ of selected ALMaQUEST galaxies to shed light on the origin of low level of star formation activity in transitioning galaxies.
 On the other hand, achieving a successful galaxy quenching model also requires a thorough  comparison between simulations and observations.

We would like to thank the anonymous referee for their careful review and valuable comments which  improved the presentation of this manuscript.
This work is supported by the National Science and Technology Council of Taiwan under grant 110-2112-M-032-020-MY3.
LL acknowledges support from NSTC 111-2112-M-001-044- and NSTC 112-2112-M-001-062-. S.F.S. thanks the PAPIIT-DGAPA AG100622 and CONACYT grant CF1939578 projects. JMS acknowledges support from the National Science Foundation under Grant No 2205551.

This paper makes use of the following ALMA data: ADS/JAO.ALMA\#2015.1.01225.S, ADS/JAO.ALMA\#2017.1.01093.S,\\ ADS/JAO.ALMA\#2018.1.00541.S, \\and ADS/JAO.ALMA\#2018.1.00558.S. ALMA is a partnership of ESO (representing its member states), NSF (USA) and NINS (Japan), together with NRC (Canada), MOST and ASIAA (Taiwan), and KASI (Republic of Korea), in cooperation with the Republic of Chile. The Joint ALMA Observatory is operated by ESO, AUI/NRAO, and NAOJ.

Funding for the Sloan Digital Sky Survey IV has been provided by the Alfred P. Sloan Foundation, the U.S. Department of Energy Office of Science, and the Participating Institutions. SDSS-IV acknowledges support and resources from the Center for High-Performance Computing at the University of Utah. The SDSS website is www.sdss.org.
SDSS-IV is managed by the Astrophysical Research Consortium for the Participating Institutions of the SDSS Collaboration including the Brazilian Participation Group, the Carnegie Institution for Science, Carnegie Mellon University, the Chilean Participation Group, the French Participation Group, Harvard-Smithsonian Center for Astrophysics, Instituto de Astrofísica de Canarias, The Johns Hopkins University, Kavli Institute for the Physics and Mathematics of the Universe (IPMU)/University of Tokyo, Lawrence Berkeley National Laboratory, Leibniz Institut f{\"u}r Astrophysik Potsdam (AIP), Max-Planck-Institut f{\"u}r Astronomie (MPIA Heidelberg), Max-Planck-Institut f{\"u}r Astrophysik (MPA Garching), Max-Planck-Institut f{\"u}r Extraterrestrische Physik (MPE), National Astronomical Observatory of China, New Mexico State University, New York University, University of Notre Dame, Observatário Nacional/MCTI, The Ohio State University, Pennsylvania State University, Shanghai Astronomical Observatory, United Kingdom Participation Group, Universidad Nacional Autónoma de México, University of Arizona, University of Colorado Boulder, University of Oxford, University of Portsmouth, University of Utah, University of Virginia, University of Washington, University of Wisconsin, Vanderbilt University, and Yale University.

\appendix
\restartappendixnumbering
\section{Radial Profiles for Individual Galaxies}
\begin{figure}
	\centering
\includegraphics[scale=0.25]{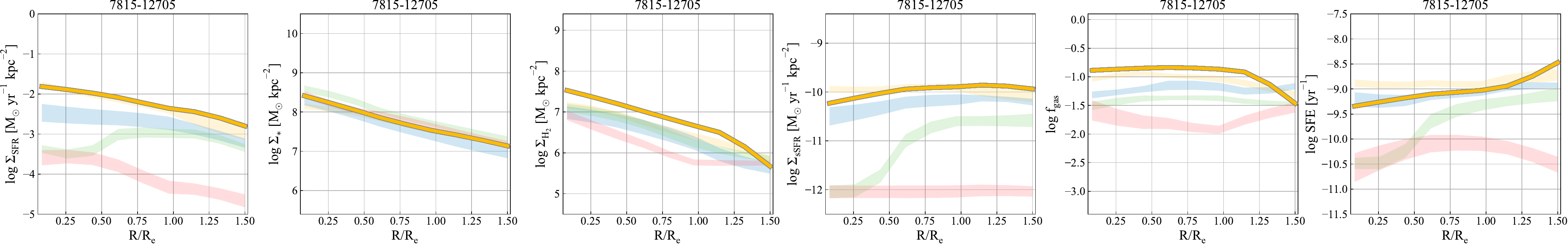} 
\includegraphics[scale=0.25]{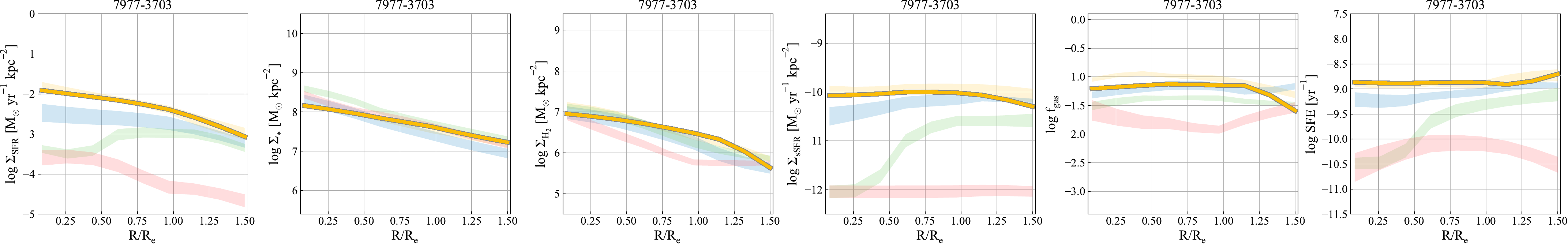} 
\includegraphics[scale=0.25]{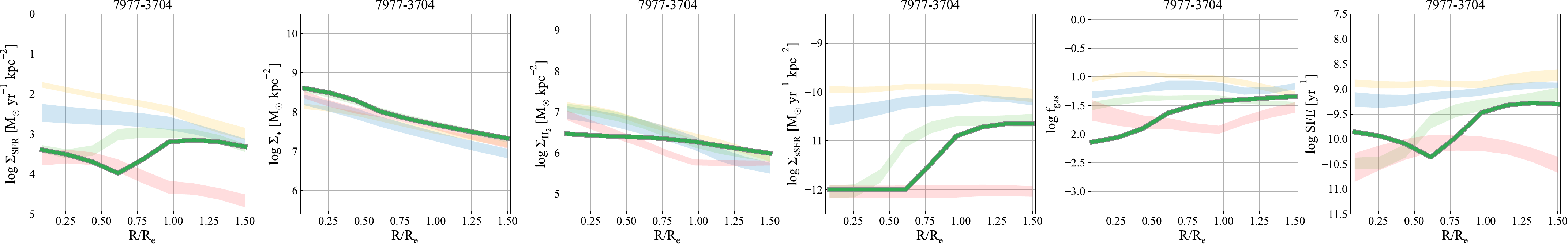} 
\includegraphics[scale=0.25]{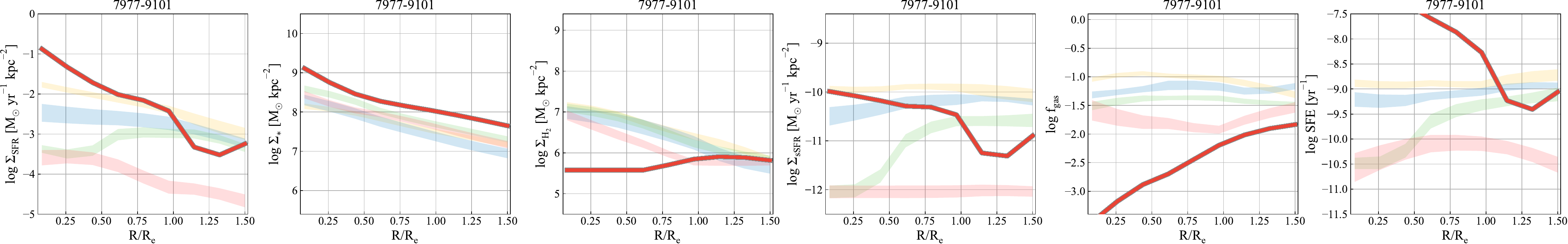} 
\includegraphics[scale=0.25]{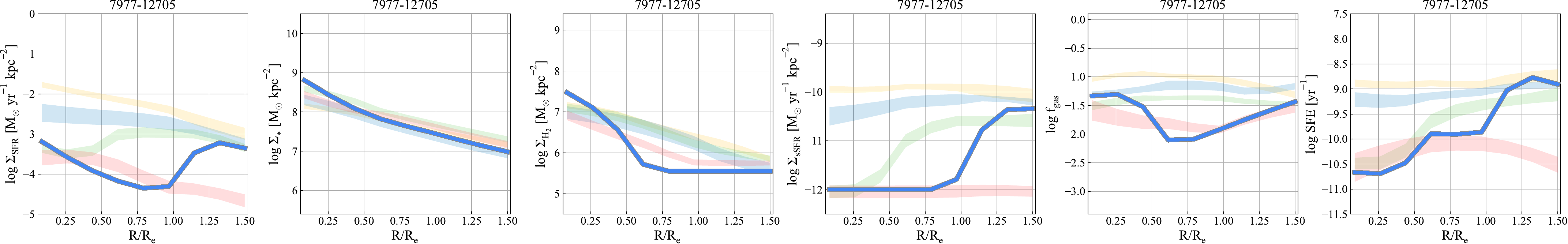} 
\includegraphics[scale=0.25]{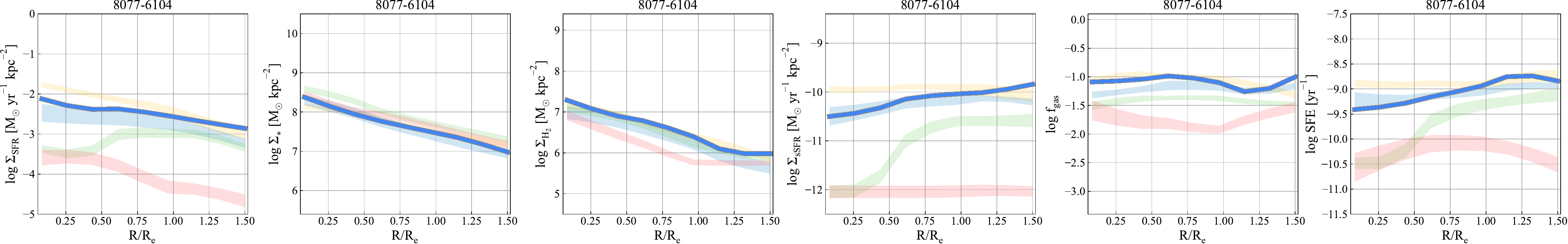} 
	\caption{Radial profiles for 46 galaxies in the ALMaQUEST main sample are presented, showing the profiles of \Sigsfr, \Sigstar, \Sigmol, \Sigssfr, \fgas, and SFE from left to right. The individual galaxy profiles are shown with opaque bands, while the transparent bands indicate the median profiles of these properties for galaxies in each category. The median profiles shown here correspond to those in Figures \ref{fig_radial_phyprops} and \ref{fig_radial_ratios}. Each line color corresponds to the galaxy category, with yellow representing cSB, blue denoting \highsf, green for \midsf, and red for \lowsf.
	}  
	\label{fig_appendix1}
\end{figure}

\begin{figure}
	\centering
\includegraphics[scale=0.25]{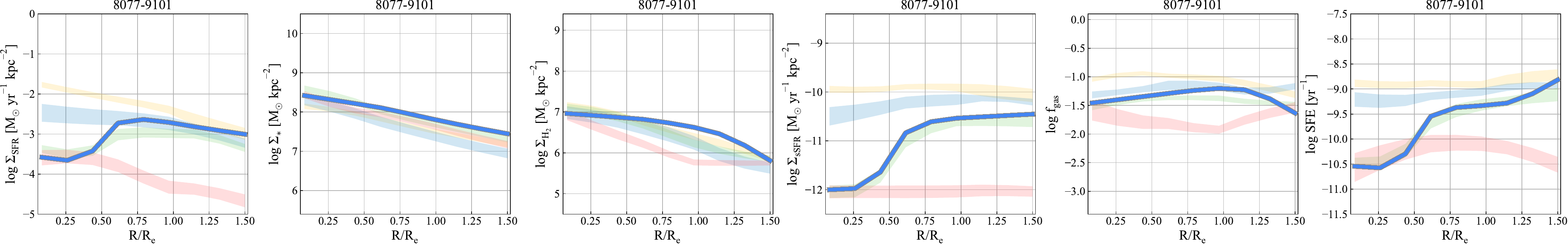} 
\includegraphics[scale=0.25]{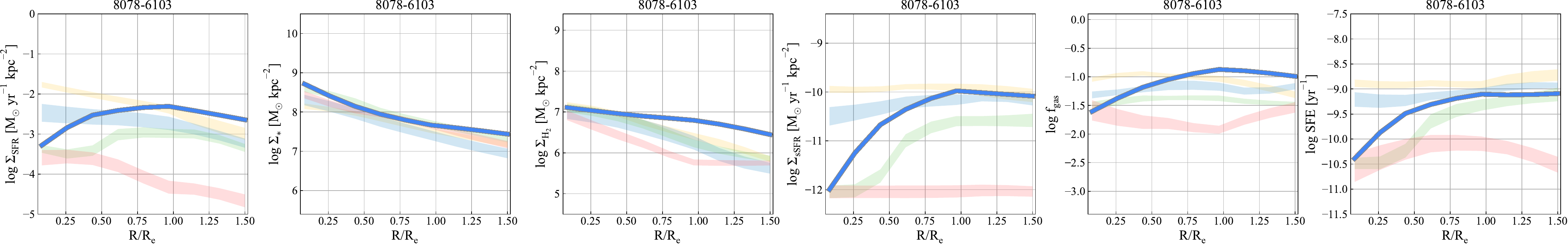} 
\includegraphics[scale=0.25]{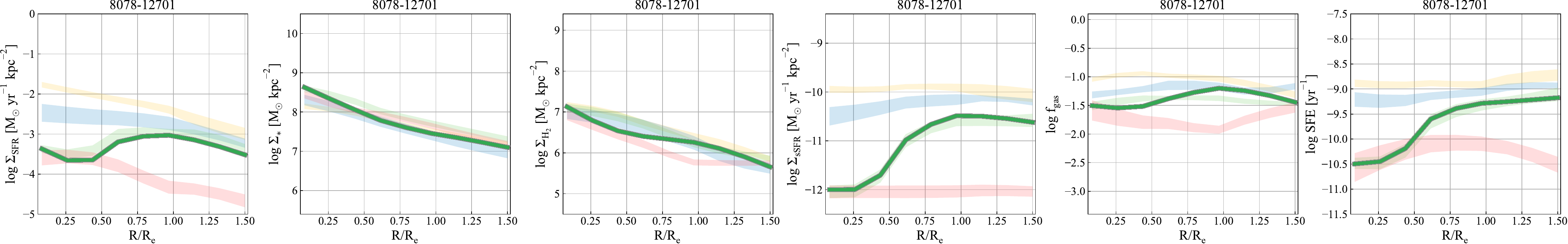} 
\includegraphics[scale=0.25]{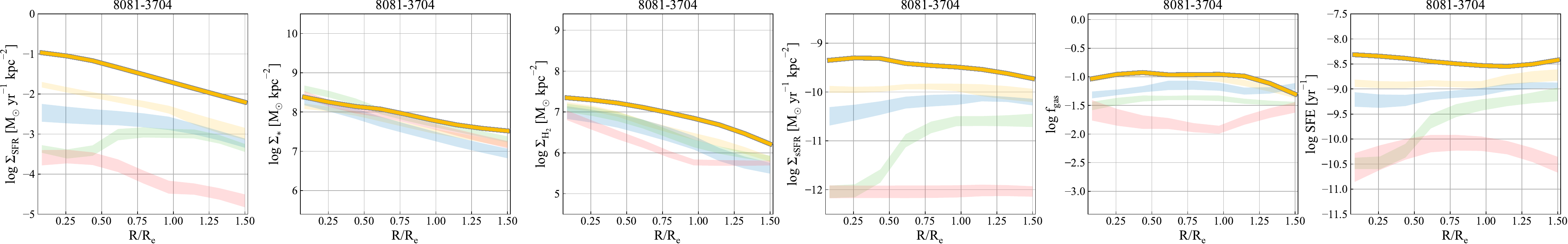} 
\includegraphics[scale=0.25]{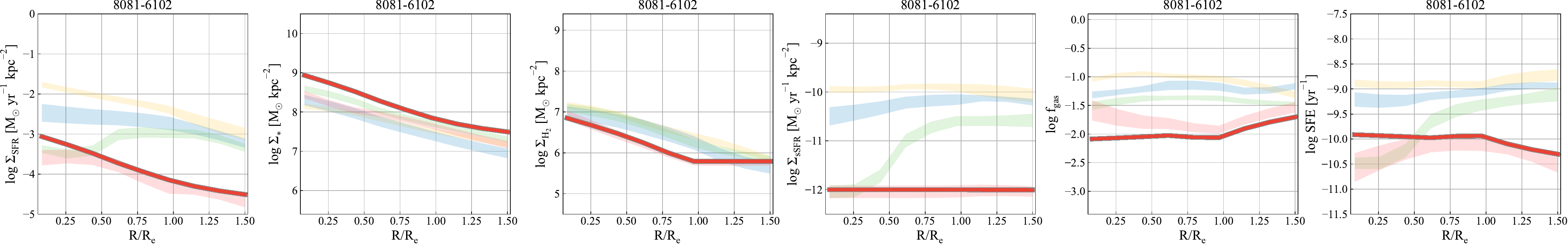} 
\includegraphics[scale=0.25]{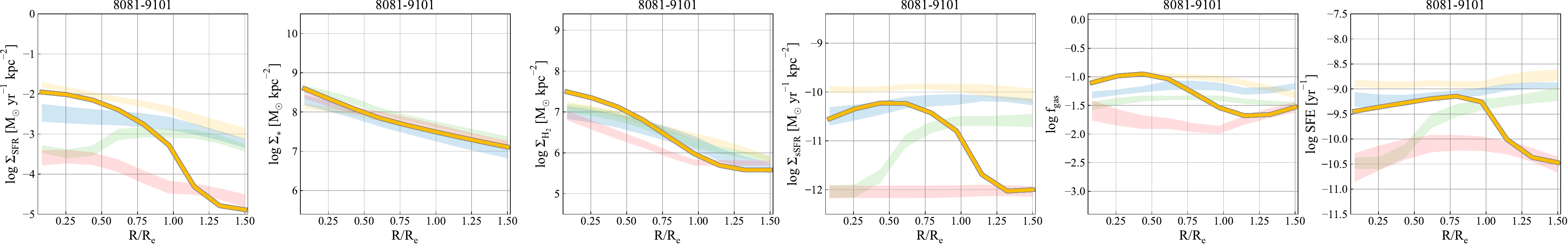} 
\includegraphics[scale=0.25]{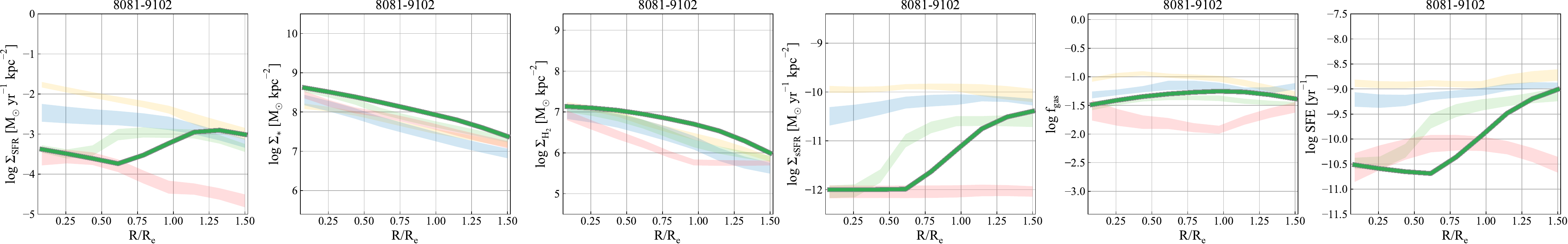} 
\includegraphics[scale=0.25]{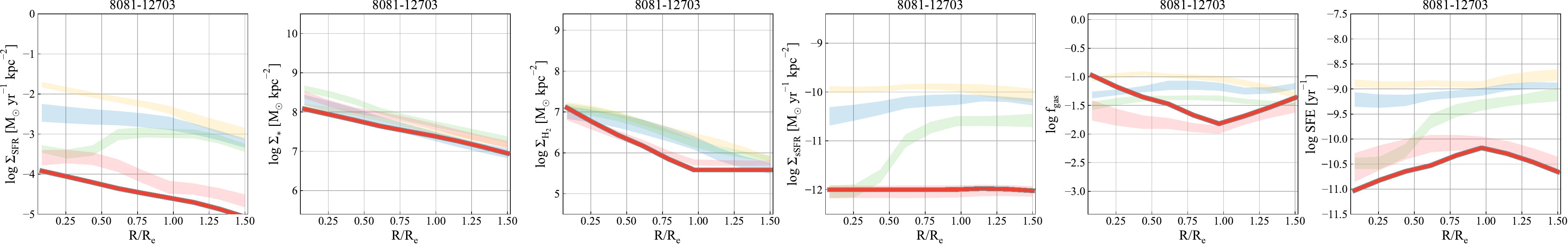} 
	\caption{Continue of Figure \ref{fig_appendix1}.
	}  
	\label{fig_appendix2}
\end{figure}

\begin{figure}
	\centering
\includegraphics[scale=0.25]{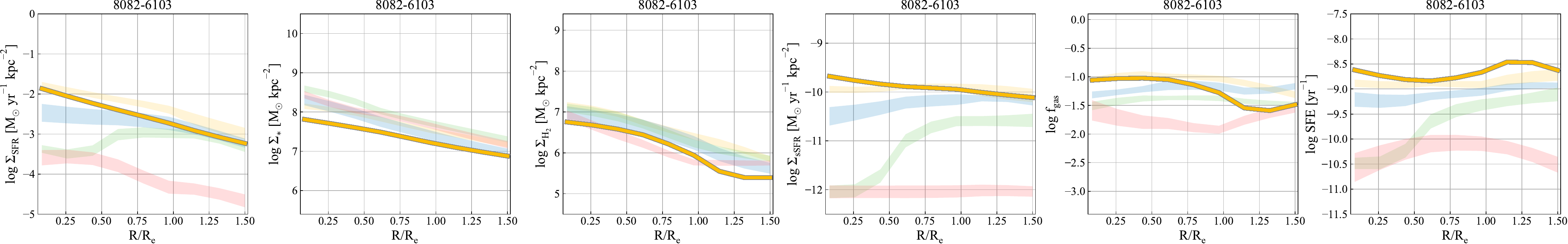} 
\includegraphics[scale=0.25]{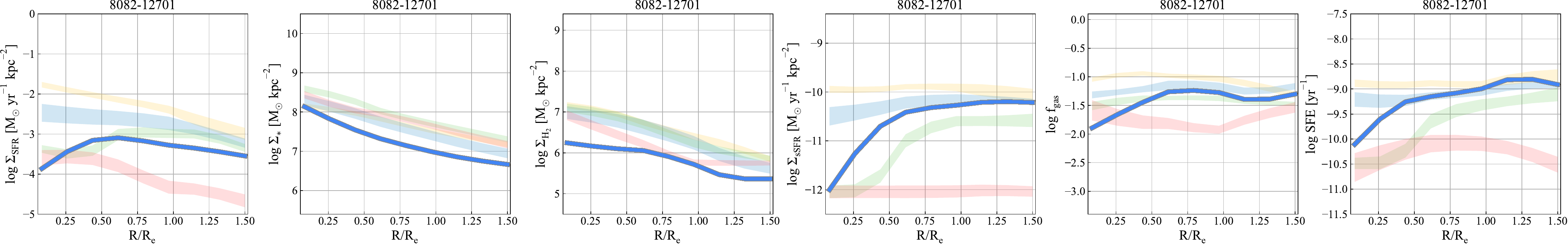} 
\includegraphics[scale=0.25]{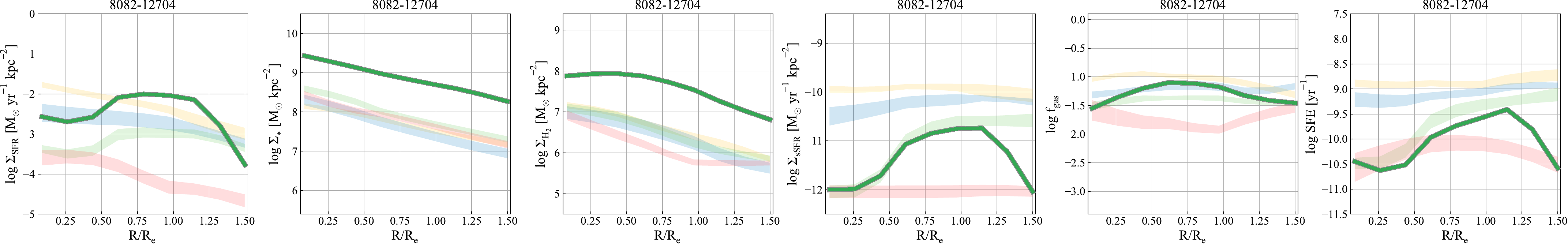} 
\includegraphics[scale=0.25]{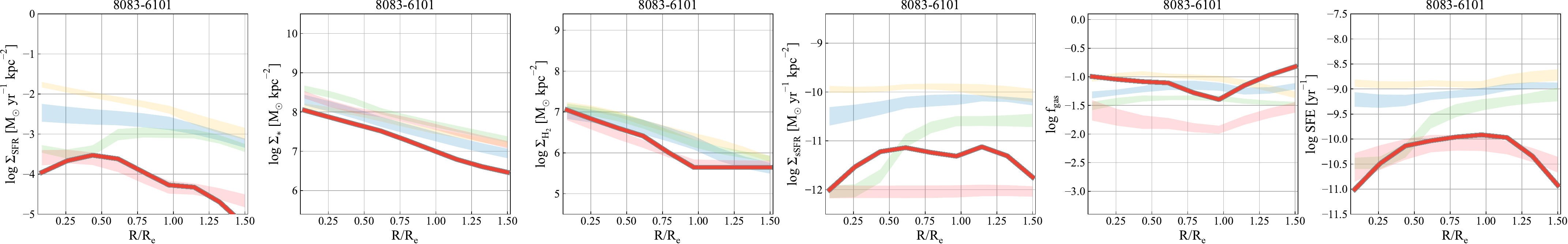} 
\includegraphics[scale=0.25]{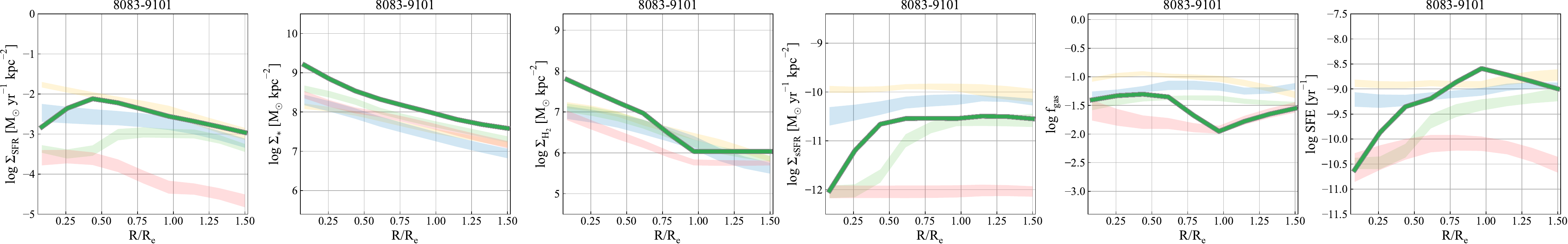} 
\includegraphics[scale=0.25]{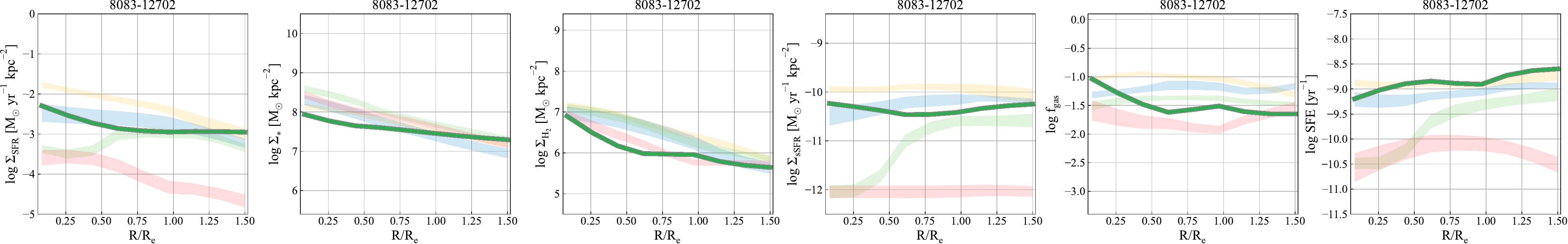} 
\includegraphics[scale=0.25]{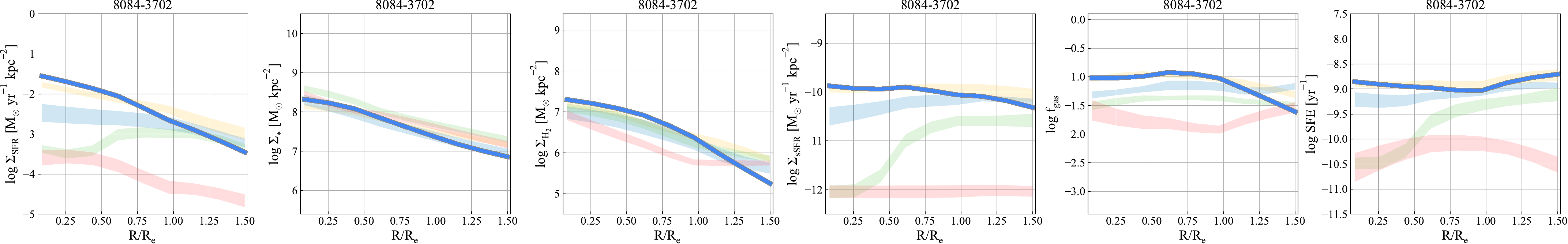} 
\includegraphics[scale=0.25]{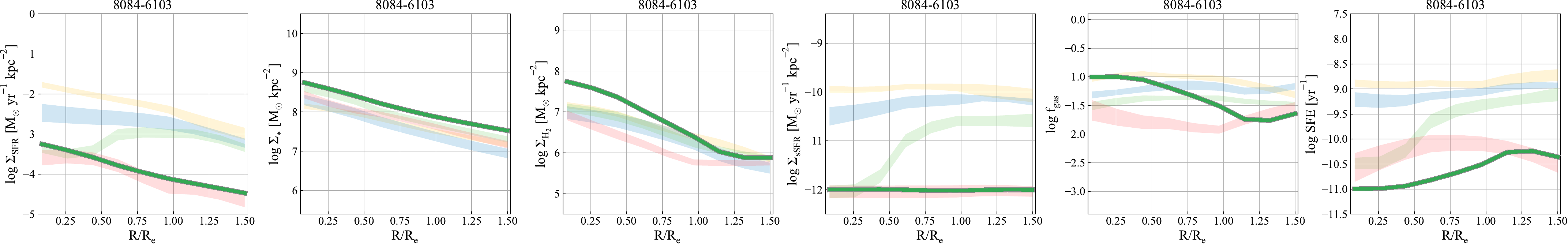} 
	\caption{Continue of Figure \ref{fig_appendix2}.
	}  
	\label{fig_appendix3}
\end{figure}

\begin{figure}
	\centering
\includegraphics[scale=0.25]{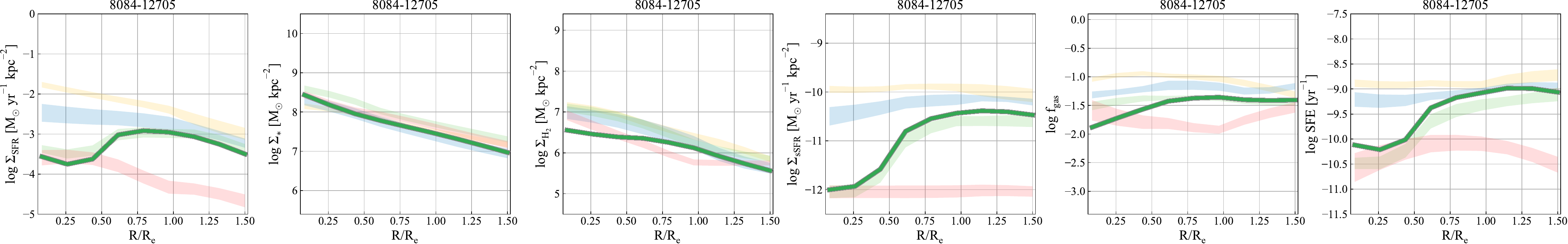} 
\includegraphics[scale=0.25]{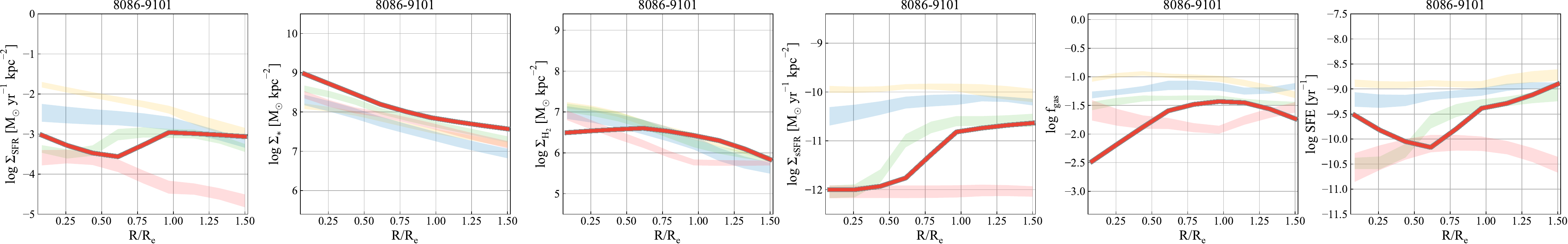} 
\includegraphics[scale=0.25]{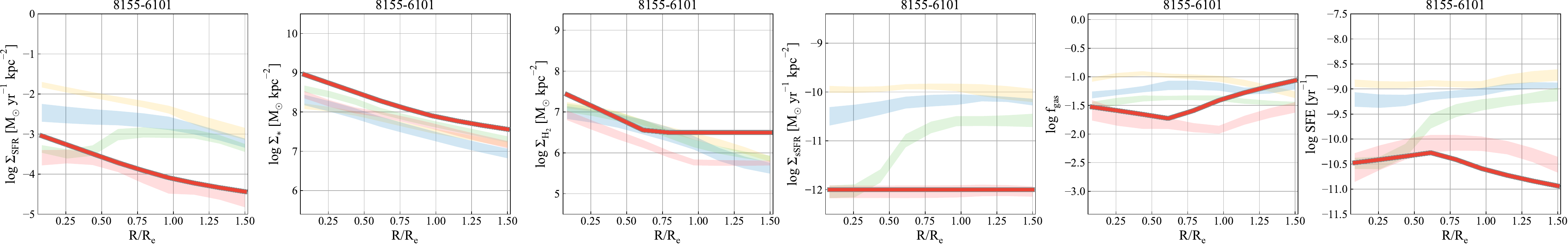} 
\includegraphics[scale=0.25]{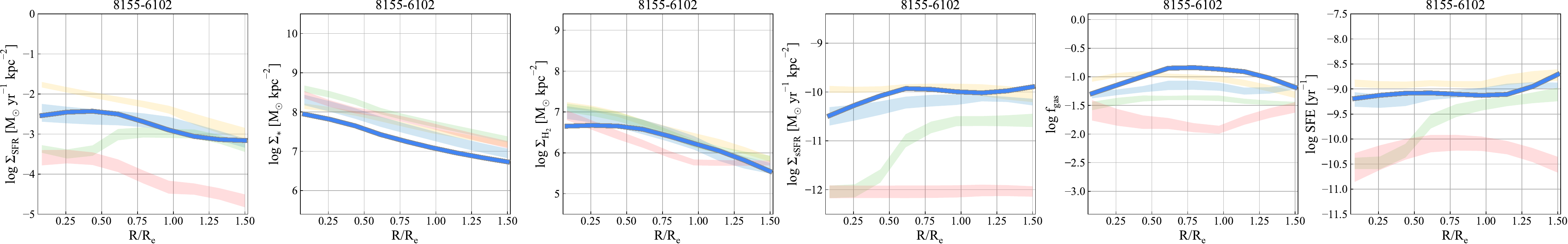} 
\includegraphics[scale=0.25]{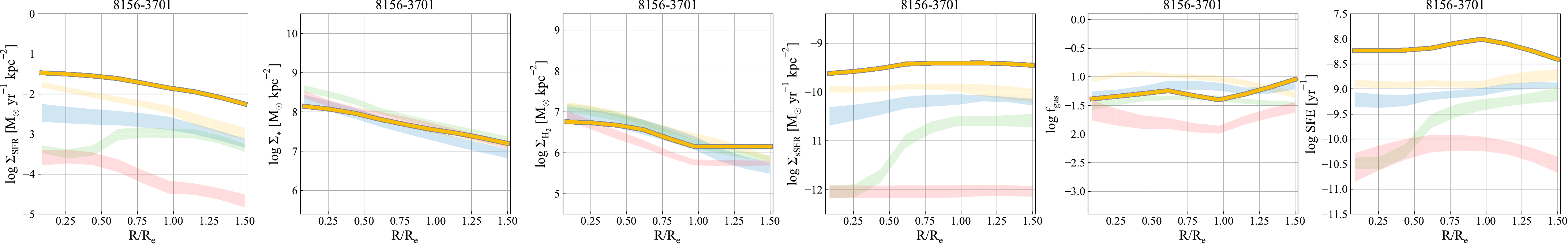} 
\includegraphics[scale=0.25]{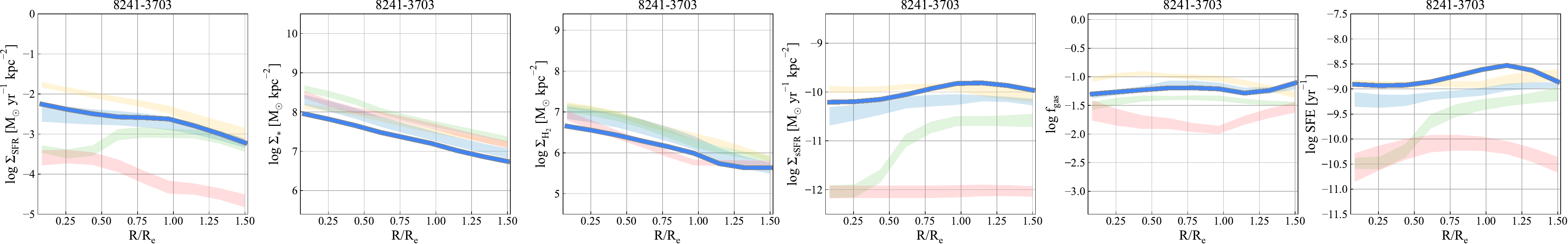} 
\includegraphics[scale=0.25]{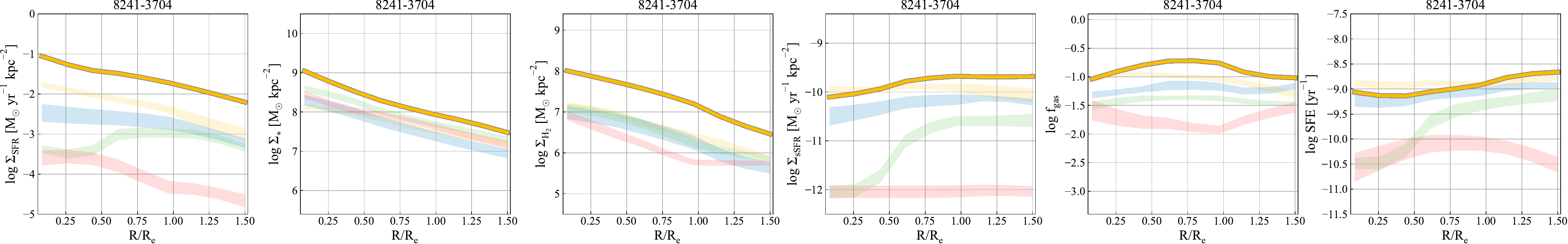} 
\includegraphics[scale=0.25]{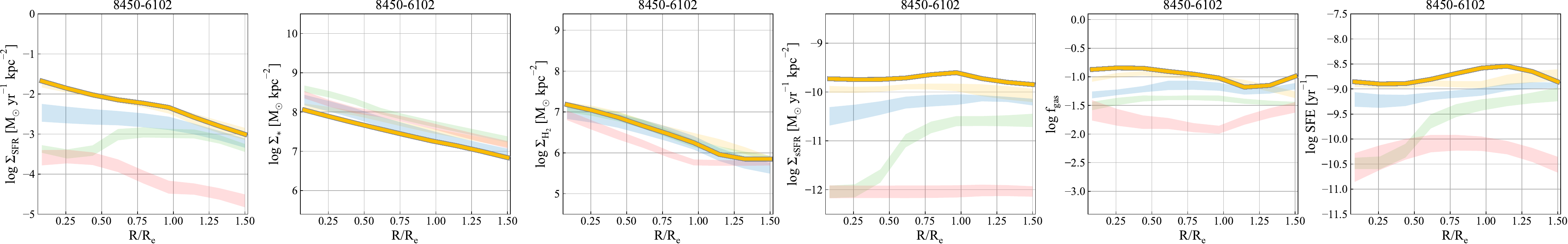} 
	\caption{Continue of Figure \ref{fig_appendix3}.
	}  
	\label{fig_appendix4}
\end{figure}

\begin{figure}
	\centering
\includegraphics[scale=0.25]{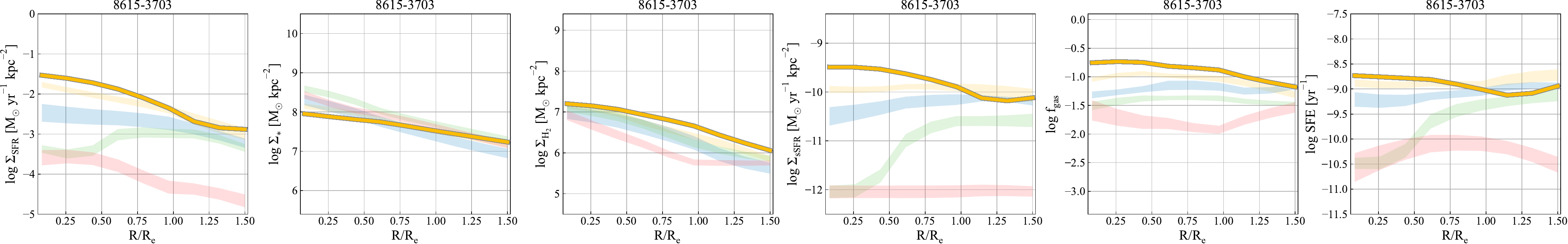} 
\includegraphics[scale=0.25]{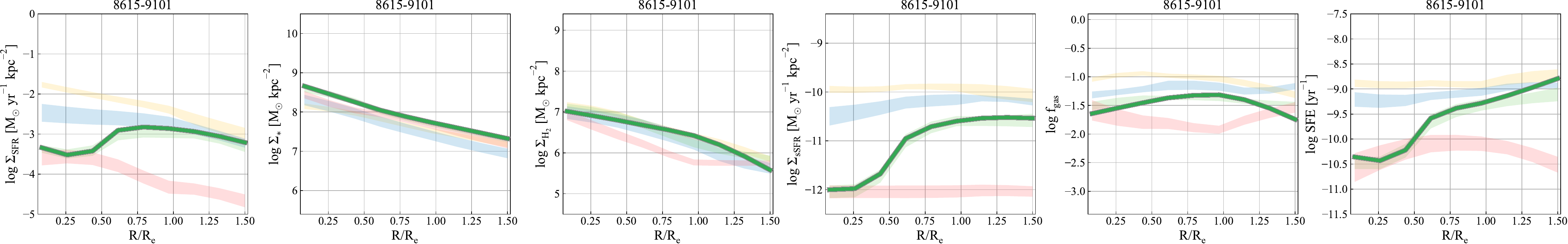} 
\includegraphics[scale=0.25]{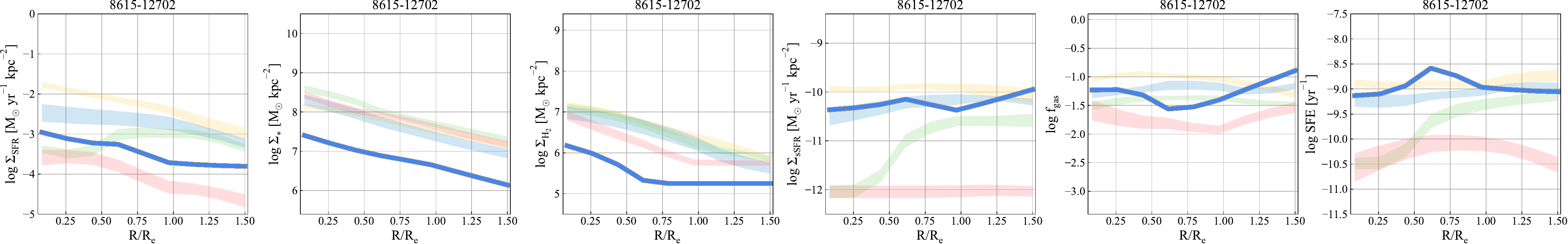} 
\includegraphics[scale=0.25]{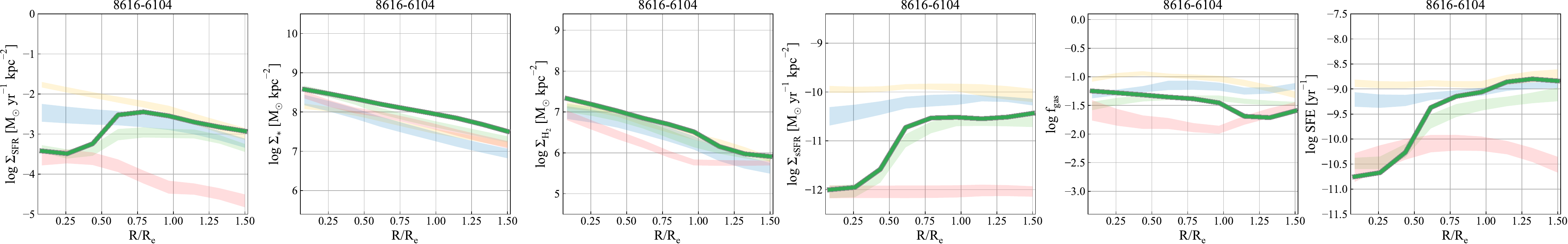} 
\includegraphics[scale=0.25]{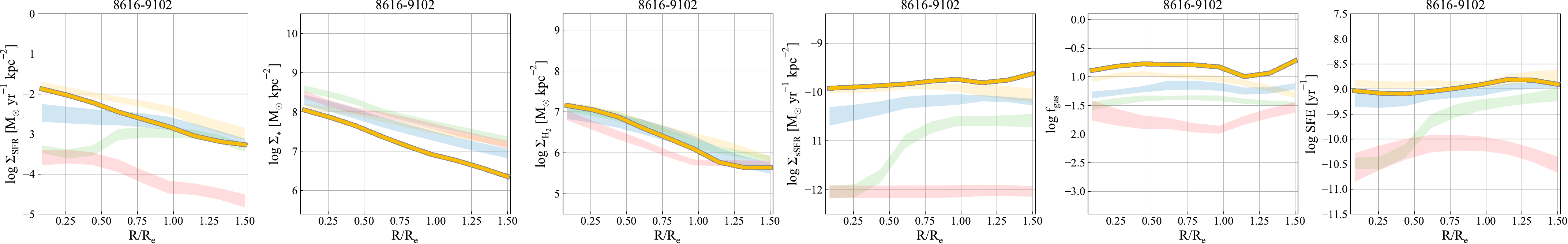} 
\includegraphics[scale=0.25]{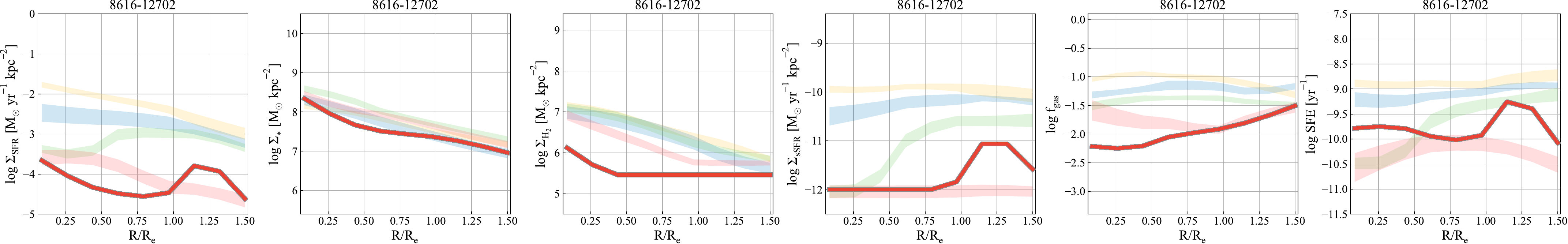} 
\includegraphics[scale=0.25]{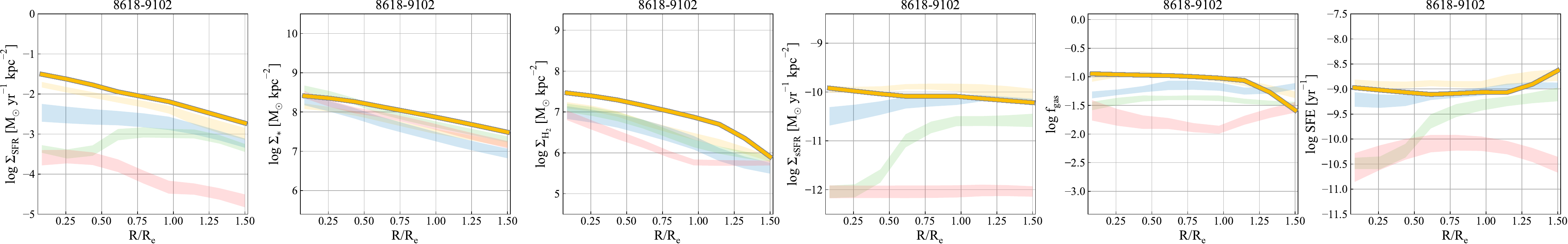} 
	\caption{Continue of Figure \ref{fig_appendix4}.
	}  
	\label{fig_appendix5}
\end{figure}

\begin{figure}
	\centering
\includegraphics[scale=0.25]{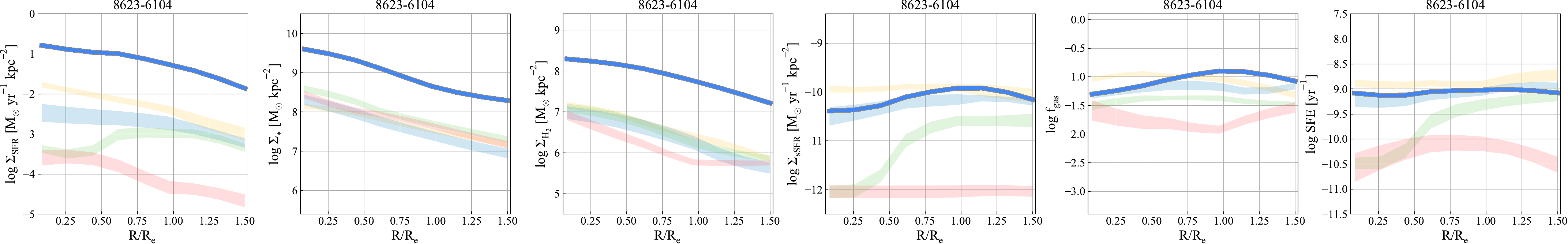} 
\includegraphics[scale=0.25]{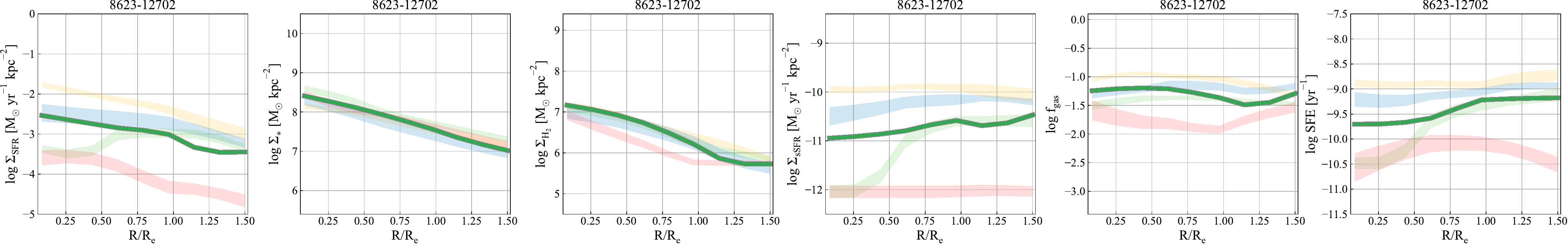} 
\includegraphics[scale=0.25]{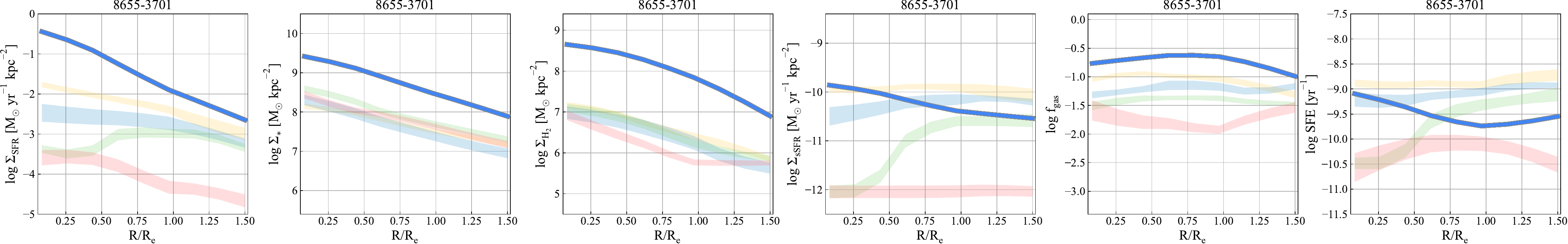} 
\includegraphics[scale=0.25]{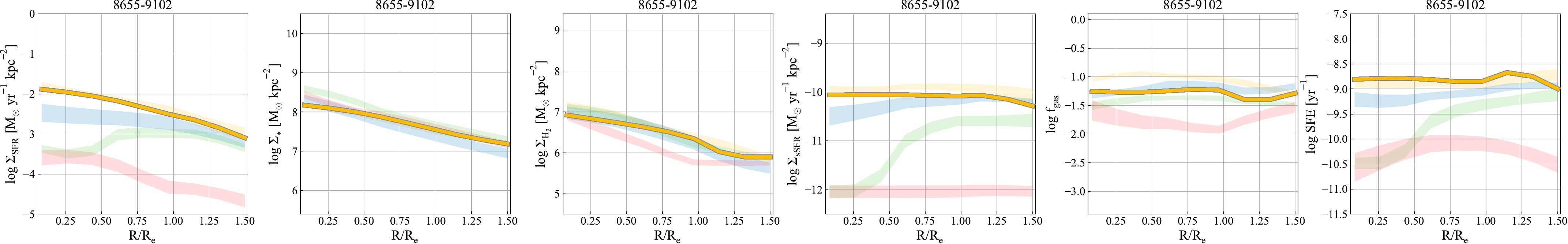} 
\includegraphics[scale=0.25]{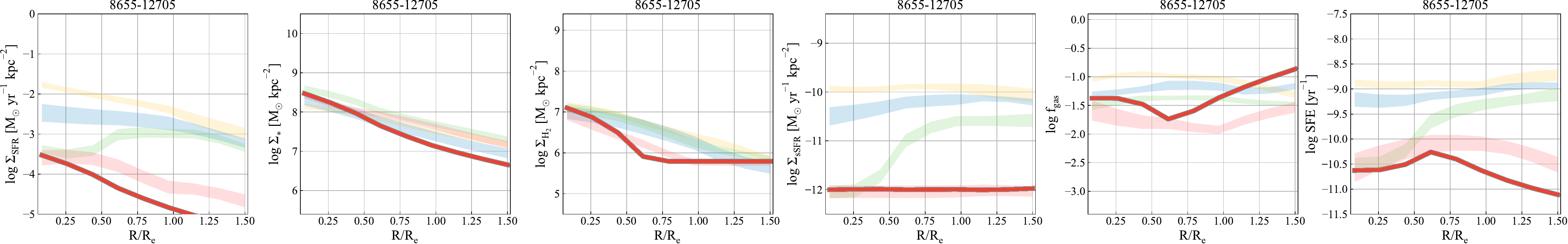} 
\includegraphics[scale=0.25]{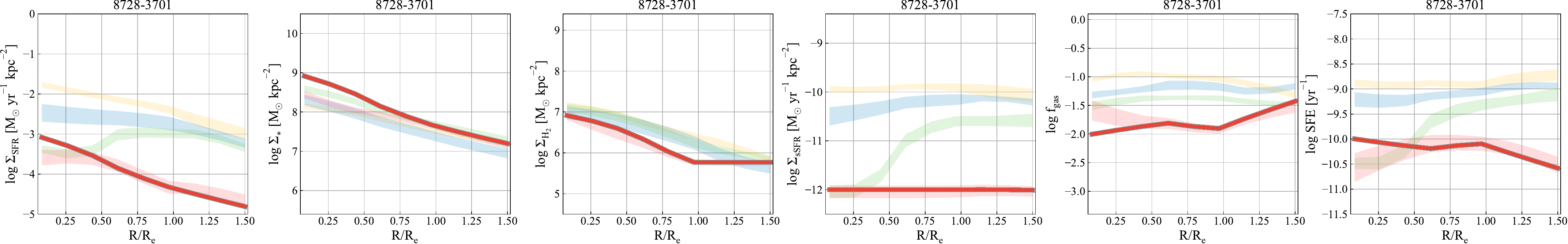} 
\includegraphics[scale=0.25]{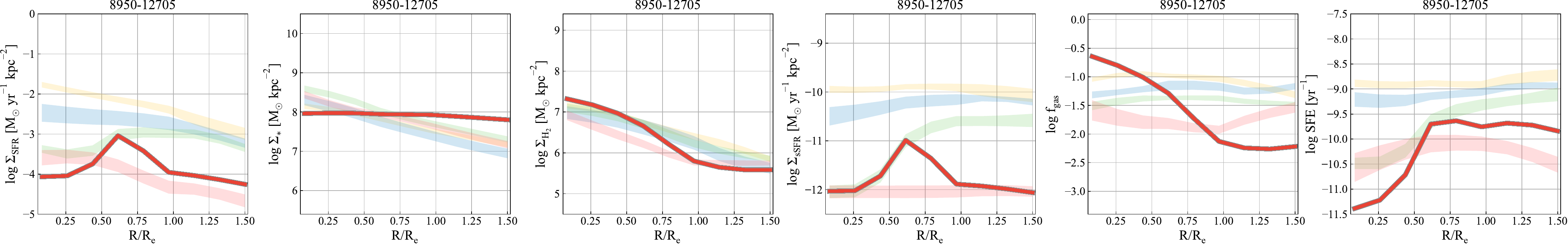} 
\includegraphics[scale=0.25]{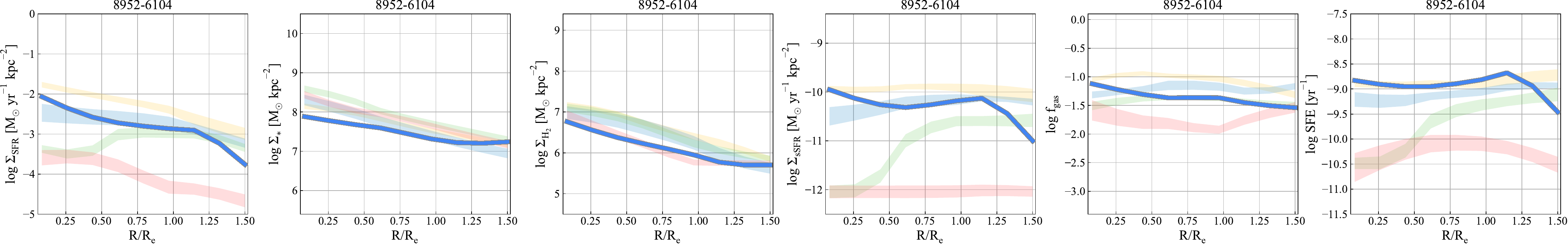} 
	\caption{Continue of Figure \ref{fig_appendix5}.
	}  
	\label{fig_appendix6}
\end{figure}

Figure \ref{fig_appendix1} to Figure \ref{fig_appendix6} displays the radial profiles of 46 galaxies in the ALMaQUEST main sample. 
For each galaxy, from left to right, the radial profiles of \Sigsfr, \Sigstar, \Sigmol, \Sigssfr, \fgas, and SFE are presented. 
Each individual galaxy's radial profile is shown as opaque dark-colored bands, with the color representing the galaxy category (yellow for cSB, blue for \highsf, green for \midsf, and red for \lowsf). 
The transparent colored bands represent the median profiles of these properties for galaxies in each category, with the thickness of the  bands indicating the standard error of the mean at each radial bin. The median profiles presented here are identical to those shown in Figures \ref{fig_radial_phyprops} and \ref{fig_radial_ratios}.

\end{CJK}
\end{document}